\documentclass[11pt,a4paper,floats,aps,nofootinbib]{revtex4-1}
\pdfoutput=1
\usepackage{latexsym}
\usepackage{amssymb}
\usepackage{epsf}
\usepackage{amsmath}
\usepackage{soul}
\usepackage{graphicx}% Include figure files

\usepackage{slashed}

\def\Slash#1{{\ooalign{\hfil$#1$\hfil\crcr\hfil$/$\hfil}}}

\begin{document}
\pagestyle{empty}

\title{Sgoldstino search at the LHC}
\author{Masaki Asano}
\email{masano@th.physik.uni-bonn.de}
\author{Raghuveer Garani}
\email{garani@th.physik.uni-bonn.de}

\affiliation{Bethe Center for Theoretical Physics \& Physikalisches Institut der 
Universit\"at Bonn, \\
Nu{\ss}allee 12, 53115 Bonn, Germany}

\begin{abstract} 
Low-scale supersymmetry breaking scenario in which the breaking scale is 
around TeV has been discussed as a possibility
to obtain a large Higgs mass  and to moderate the fine tuning problem. 
A characteristic feature is that the hidden sector would be accessible at colliders in such a scenario.
In this paper, we investigate the phenomenology of sgoldstino which is the scalar component of the goldstino superfield. 
We present partial widths and branching ratios for sgoldstinos decaying to final states involving Higgs bosons and sparticles which have not been discussed in detail so far.

\end{abstract} 
\maketitle
%%%%%%%%%%%%%%%%%%%%%%%%%%%%%%%%%%%%%%%%%%%%%%%%%%%%%%%%%%%%%%%%%%%%%%%%%%%%
%\newpage
\baselineskip=18pt
\setcounter{page}{2}
\pagestyle{plain}
\baselineskip=18pt
\pagestyle{plain}

\setcounter{footnote}{0}

%%%%%%%%%%%%%%%%%%%%%%%%%%%%%%%%%%%%%%%%%%%%%%%%%%%%%%%%%%%%%%%%%%%%%%%%%%%% 
% Introduction 
%%%%%%%%%%%%%%%%%%%%%%%%%%%%%%%%%%%%%%%%%%%%%%%%%%%%%%%%%%%%%%%%%%%%%%%%%%%% 
\section{Introduction}

Supersymmetry (SUSY) is an interesting possibility to explain the smallness of the electroweak symmetry breaking scale. In SUSY, the electroweak symmetry breaking scale can be interpreted in terms of soft breaking parameters (and $\mu$ parameter) thus SUSY particles are plausible candidates for new particles that can be produced at the LHC. 

Phenomenology of the minimal supersymmetric standard model (MSSM) has widely been studied. There are several possibilities of mediation schemes of SUSY breaking, however, only MSSM particles can be accessible by current colliders in many scenarios \footnote{
One of the exceptions is the case of gravitino lightest superpartner particle (LSP). For example, in gauge mediation the next-LSP will decay to gravitino before exiting the detector in some region of the parameter space. 
}.
Since the mediation scale is much higher than the electroweak scale, other sectors are decoupled.

%%%%%%%%%%%%%%%%%%%%%%%%%%%%%%%%%%%%%%%%%%%%%%%%%%%%%%%%%%%%%%%%%%
On the other hand, considering very low scale mediation and low scale SUSY breaking $\sim {\mathcal O} (1)$ TeV is still possible 
%%%%%%%%%
\cite{Brignole:1996fn,Brignole:2003cm,Itoh:2006fv,Dine:2007xi,Antoniadis:2008es,Antoniadis:2009rn,Antoniadis:2010hs,Murayama:2012jh}.
In this case, couplings with the hidden sector is not strongly suppressed and consequently affects collider phenomenology. For example, it is possible to produce sgoldstino which is the scalar superpartner of goldstino 
%%%%%%%%%
\cite{Perazzi:2000id,Perazzi:2000ty,Gorbunov:2000ht,Gorbunov:2001pd,Gorbunov:2002er,Demidov:2004qt,Bellazzini:2012mh,Petersson:2012nv,Dudas:2012fa,Astapov:2014mea,Bhattacharya:1988ey}
. Furthermore, higher dimensional operators in such a scenario can affect the lightest Higgs boson mass~
%%%%%%%%%
\cite{Polonsky:2000rs,Casas:2003jx,Dine:2007xi,Cassel:2009ps,Carena:2009gx,Antoniadis:2009rn,Cassel:2010px,Carena:2010cs,Antoniadis:2010hs,Antoniadis:2010nb,Cassel:2011zd,Petersson:2011in,Boudjema:2012cq} and its impact on naturalness is discussed in 
%%%%%%%%%%%
\cite{Casas:2003jx,Cassel:2009ps,Cassel:2010px,Antoniadis:2010nb,Cassel:2011zd,Antoniadis:2014eta}.
%%%%%%%%%%%%%%%%%%%%%%%%%%%%%%%%%%%%%%%%%%%%%%%%%%%%%%%%%%%%%%%%%%

In this paper, we investigate low-scale SUSY breaking scenario and specifically study the collider phenomenology of sgoldstino.
We present the branching ratios of sgoldstino to Higgs boson final states and SUSY particle final states which have not been studied in detail so far. 
The decay to Higgs bosons is induced, for example, by $(\mu B\mu/F) \phi_x |H_u|^2$ term (for details, see Section 5). Since this term is not proportional to the electroweak vacuum expectation value (VEV), this decay mode can be important. 
As one can expect from the equivalence theorem, we also show that the branching ratios to the longitudinal mode of weak gauge bosons are similar to that of the Higgs branch in heavy sgoldstino parameter region.\footnote{
The branching ratios to the longitudinal mode of weak bosons have been studied, for example, in Refs.~\cite{Bellazzini:2012mh,Astapov:2014mea}. 
}

The remainder of this paper is organized as follows. In the next section, we introduce a simple effective Lagrangian as an example model of low-scale SUSY breaking scenario. We present Higgs-sgoldstino potential and the Higgs-sgoldstino mixing in Section 3 and the Gaugino-Higgsino-Goldstino mass matrix in Section 4. Then, we study sgoldstino production at the LHC and their subsequent decays in Section 5 and Section 6 is devoted to the summary. 

%\newpage

%%%%%%%%%%%%%%%%%%%%%%%%%%%%%%%%%%%%%%%%%%%%%%%%%%%%%%%%%%%%%%%%%%%%%%%%%%%% 
%  Model 
%%%%%%%%%%%%%%%%%%%%%%%%%%%%%%%%%%%%%%%%%%%%%%%%%%%%%%%%%%%%%%%%%%%%%%%%%%%% 
\section{Lagrangian}
We study the phenomenology of sgoldstino in a simple model which includes MSSM superfields and a singlet sgoldstino chiral superfield $X = \phi_X + \sqrt{2} \theta \psi_X + \theta^2 F_X$. The auxiliary component $F_X$ has a non-zero VEV. The fermionic component $\psi_X$ corresponds to goldstino and the scalar component $\phi_X$ correspond to scalar and pseudo-scalar boson called sgoldstino and pseudo-sgoldstino, respectively. We consider the following simple lagrangian $\mathcal{L}_X$, 
%--------------------------->
\begin{eqnarray}
 \mathcal{L}_X &=& 
\int d\theta^4 \left( 1 - \frac{1}{4}\frac{m_X^2}{F^2} X^\dagger X \right) X^\dagger X 
   + \left( \int d\theta^2 F X + h.c.  \right), 
  \label{eq:LagrangianX}
\end{eqnarray}
%---------------------------<
where the non-zero F-term VEV is $\langle F_X \rangle = -F$ and masses of sgoldstino and pseudo-sgoldstino are obtained to be $m_X$. 

In addition to Eq.~\eqref{eq:LagrangianX}, we consider the following usual MSSM sector 
in the lagrangian $\mathcal{L} = \mathcal{L}_K + \mathcal{L}_W + \mathcal{L}_X$, 
%--------------------------->
\begin{eqnarray}
 \mathcal{L}_K &=& 
\int d\theta^4 \left[ 
  \left( 1 - \frac{m_{\tilde{f}_i}^2}{F^2} X^\dagger X \right) \Phi_i^\dagger e^V \Phi_i
 +\left( 1 - \frac{m_{H_{u,d}}^2}{F^2} X^\dagger X \right) H_{u,d}^\dagger e^V H_{u,d}
\right . \nonumber \\ 
&& \left . \qquad \qquad + \left\{
 -\left( \frac{\mu}{F} X^\dagger + \frac{B_\mu}{F^2} X^\dagger X \right) H_d \cdot H_u
   + h.c. \right\} \right],
\nonumber \\
 \mathcal{L}_W &=& 
\int d\theta^2 \left[ 
    \frac{1}{4} \left( 1 + \frac{2 M_a}{F} X \right) {\bf Tr}[W^{a \alpha} W^a_\alpha] 
  + \left( y_e + \frac{A_e}{F} X \right) H_d \cdot L E^c 
\right . \nonumber \\ 
&& \left . \qquad \qquad 
  + \left( y_d + \frac{A_d}{F} X \right) H_d \cdot Q D^c 
  + \left( y_u + \frac{A_u}{F} X \right) H_u \cdot Q U^c 
              \right]
  + h.c. , 
  \label{eq:Lagrangian0}
\end{eqnarray}
%---------------------------<
where, $\alpha \cdot \beta = \epsilon_{ij} \alpha^i \beta^j$ and $\epsilon_{12}=1$. For simplicity, we assume all soft SUSY breaking parameters and $\mu$ term are real.

General lagrangian for low-scale SUSY breaking scenario consists of many more possible operators as discussed in~\cite{Dudas:2012fa}.
However, this simple lagrangian would be adequate to investigate the phenomenology of sgoldstino at colliders. For example, there is no difference when we consider the $\mu$ and $B_\mu$ terms to originate from $\mathcal{L}_W \supset \mu_w H_d \cdot H_u + (B_{\mu w}/F) X H_d \cdot H_u$ instead of $\mu$ and $B_\mu$ terms presented in Eq.~\eqref{eq:Lagrangian0}, up to $\mathcal{O}(1/F)$. Although, the  term $\mathcal{L}_W \supset (A_X/F) X X H_d \cdot H_u$ can alter the sgoldstino phenomenology if it exists, therefore we assume these are small for simplicity. The full lagrangian up to $\mathcal{O}(1/F^2)$ is presented in Appendix D.

As will be shown in Section 5, sgoldstino decays via $\mathcal{O}(1/F)$ suppressed couplings. In this paper, we investigate the phenomenology at the leading order and neglect $\mathcal{O}(1/F^2)$ and higher order terms \footnote{Except in the numerical calculation of neutral higgs masses, see Section 3.2 for details.}. If $m_{\rm soft}^2/F$ is not small, the expansion does not work, resulting in higher dimensional operators becoming non-negligible. Thus, for predictability of this effective Lagrangian, we only consider the parameter space in which $m_{\rm soft} < \sqrt{F}$.

%%%%%%%%%%%%%%%%%%%% 
%  potential 
%%%%%%%%%%%%%%%%%%%%
\section{Higgs-sgoldstino potential}

In this section we start with the presentation of Higgs and sgoldstino potential for this model. 
Electroweak symmetry breaking causes Higgs-sgoldstino mixing (and pseudo-Higgs - pseudo-sgoldstino mixing). We solve for the minimization conditions and define mass eigenbasis for such scalar fields. 

\subsection{Potential} 
The Higgs-sgoldstino potential is provided by D- and F-terms contributions, $V_{h-s} = V_D + V_F$, where   
%--------------------------->
\begin{eqnarray}
 V_D &=& \frac{g^{\prime 2}}{8} \left( 1 + \frac{2 M_1}{F} \frac{\phi_X + \phi_X^*}{2} \right)^{-1}
         \left(  |H_u|^2 - |H_d|^2 \right)^2 
\\ &&   
        +\frac{g_2^2}{8} \left( 1 + \frac{2 M_2}{F} \frac{\phi_X + \phi_X^*}{2} \right)^{-1}
         \left(  H_u^\dagger \sigma^i H_u
                + H_d^\dagger \sigma^i H_d
        \right)^2, 
\nonumber 
  \label{eq:VD}
\\
 V_F &=& \left|- \left( \mu + \frac{B_\mu}{F} \phi_X \right) \epsilon_{ij}H_d^i 
               +\frac{m_{H_u}^2}{F} \phi_X H_u^{*j}
         \right|^2
        +\left|- \left( \mu + \frac{B_\mu}{F} \phi_X \right)\epsilon_{ij} H_u^j 
               +\frac{m_{H_d}^2}{F} \phi_X H_d^{*i}
         \right|^2
\nonumber \\ && 
        + m_X^2|\phi_X|^2
        + m_{H_u}^2 |H_u|^2 
        + m_{H_d}^2 |H_d|^2 
        + B_\mu \left\{ H_d \cdot H_u + (H_d \cdot H_u)^\dagger \right\}
, 
  \label{eq:VF}
\end{eqnarray}
%---------------------------<
up to $\mathcal{O}(1/F)$. For $\mathcal{O}(1/F^2)$ terms, see Appendix D. 
Note that, we write the scalar components of up-type and down-type Higgs, $H_u$ and $H_d$, by the same characters as that of the superfields. The vacuum expectation values $v_d$, $v_u$ and $v_X$ are defined as 
%--------------------------->
\begin{eqnarray}
 H_d &=& \left(
\begin{array}{cc}
      ( h_d^0 + i A_d + v_d )/\sqrt{2}\\
        H_d^-     \\ 
\end{array}
       \right),
\quad 
 H_u  = \left(
\begin{array}{cc}
       H_u^+    \\
      ( h_u^0 + i A_u + v_u )/\sqrt{2}\\ 
\end{array}
       \right),
\nonumber \\ 
 \phi_X  &=& \left( s_X + i a_X + v_X \right)/\sqrt{2},
  \label{eq:Higgsfield}
\end{eqnarray}
%---------------------------<
where $v^2 =  v_d^2 + v_u^2 \sim (246 {\rm GeV})^2$ and we define $\tan\beta = v_u/v_d$.

The vacuum conditions, up to $\mathcal{O}(1/F)$, are obtained as  
%--------------------------->
\begin{eqnarray}
 \frac{1}{2}m_Z^2 &=& - \mu^2 + \frac{ m_{H_d}^2 - m_{H_u}^2 \tan^2\beta }{\tan^2\beta - 1},
  \label{eq:vx} \\ 
 \sin2\beta &=&  - \frac{2 B_\mu}{m_A^2},
\nonumber \\
 v_X &=& -\frac{1}{2\sqrt{2}F} \frac{v^2}{m_X^2} 
          \left[ 
             \left( m_A^2 - 2 \mu^2 \right) \mu \sin2\beta 
           +2\mu B_\mu 
           -\frac{(\cos2\beta)^2}{2 c_W^2} m_W^2 \left( s_W^2 M_1 + c_W^2 M_2 \right) 
         \right], 
\nonumber 
\end{eqnarray}
%---------------------------<
the definition of $m_A^2$ is the same as that of the usual MSSM, $m_A^2 = m_{H_d}^2 + m_{H_u}^2 + 2\mu^2$. As it can be seen in Eq.~\eqref{eq:vx}, neglecting $\mathcal{O}(1/F^2)$ and further higher order terms results in the first two conditions being the same as that of MSSM.  We can neglect $v_X$ hereafter since it is $1/F$ suppressed and all terms which accompany $v_X$ are further suppressed by factor $1/F$.

\subsection{Neutral scalar mass matrix}

The neutral scalar mass terms are written as 
%--------------------------->
\begin{eqnarray}
&& \mathcal{L} \supset - \frac{1}{2}
\left( 
h_u^0 ~ h_d^0 ~ s_X 
\right)
\left( 
\begin{array}{ccc}
       m_A^2 \cos^2\beta + m_Z^2 \sin^2\beta &  -(m_A^2 + m_Z^2) \cos\beta \sin\beta   &  ({\rm m}_{\rm higgs^0}^2)_{13} \\
       -(m_A^2 + m_Z^2) \cos\beta \sin\beta &   m_A^2 \sin^2\beta + m_Z^2 \cos^2\beta  & ({\rm m}_{\rm higgs^0}^2)_{23} \\ 
       ({\rm m}_{\rm higgs^0}^2)_{31}   & ({\rm m}_{\rm higgs^0}^2)_{32} & m_X^2\\ 
\end{array}
\right)
\left(
\begin{array}{c}
       h_u^0 \\
       h_d^0 \\ 
       s_X   \\ 
\end{array}
\right), ~~~
\nonumber \\ 
&&
({\rm m}_{\rm higgs^0}^2)_{13} = ({\rm m}_{\rm higgs^0}^2)_{31} 
= \frac{ v }{ \sqrt{2} F } \left[ ( -2 \mu^2 + m_A^2 \cos2\beta ) \mu \cos\beta 
                                   +m_Z^2 (s_W^2 M_1 + c_W^2 M_2) \cos2\beta \sin\beta                       \right],
\nonumber \\ &&
({\rm m}_{\rm higgs^0}^2)_{23} = ({\rm m}_{\rm higgs^0}^2)_{32} 
= -\frac{v}{\sqrt{2} F}
\left[ ( 2 \mu^2 + m_A^2 \cos2\beta ) \mu \sin\beta 
+ m_Z^2 (s_W^2 M_1 + c_W^2 M_2) \cos2\beta \cos\beta \right],
\nonumber \\
  \label{eq:HiggsMass0}
\end{eqnarray}
%---------------------------<
up to $\mathcal{O}(1/F)$. By the usual MSSM rotation, 
%--------------------------->
\begin{eqnarray}
\left(
\begin{array}{c}
        h\\
        H\\ 
\end{array}
\right)
&=& 
\left(
\begin{array}{cc}
       \cos\alpha & -\sin\alpha \\ 
       \sin\alpha & \cos\alpha \\ 
\end{array}
\right)
\left(
\begin{array}{c}
        h_u^0\\
        h_d^0\\ 
\end{array}
\right), \qquad
\tan2\alpha = \tan2\beta \frac{m_A^2+m_Z^2}{m_A^2-m_Z^2},
  \label{eq:MSSMHiggs1}
\end{eqnarray}
%---------------------------<
Eq.~\eqref{eq:HiggsMass0} is rewritten as
%--------------------------->
\begin{eqnarray}
 \mathcal{L} &\supset& - \frac{1}{2}
\left( 
h ~H ~ s_X 
\right)
\left( 
\begin{array}{ccc}
       m_h^2 &       0  &  ({\rm m}_{\rm higgs^0}^2)_{13}^\prime \\
          0  &   m_H^2  & ({\rm m}_{\rm higgs^0}^2)_{23}^\prime \\ 
       ({\rm m}_{\rm higgs^0}^2)_{31}^\prime   & ({\rm m}_{\rm higgs^0}^2)_{32}^\prime & m_X^2\\ 
\end{array}
\right)
\left(
\begin{array}{c}
       h \\
       H \\ 
       s_X   \\ 
\end{array}
\right),
  \label{eq:HiggsMass1}
\end{eqnarray}
%---------------------------<
%--------------------------->
\begin{eqnarray}
m_{h,H}^2 &=& \frac{1}{2}(m_Z^2 + m_A^2) 
             \mp \frac{1}{2} \sqrt{(m_Z^2-m_A^2)^2 + 4 m_Z^2 m_A^2 (\sin2\beta)^2}. 
  \label{eq:HiggsMass1_add}
\end{eqnarray}
%---------------------------<
%
In the limit $m_A \gg m_Z$,  $\sin2\alpha = - \sin2\beta$,  the off-diagonal components can be written as 
%--------------------------->
\begin{eqnarray}
({\rm m}_{\rm higgs^0}^2)_{13}^\prime &=& ({\rm m}_{\rm higgs^0}^2)_{31}^\prime 
= \frac{ v }{ \sqrt{2} F } 
    \left[ 2 \mu^3 \sin2\beta + m_Z^2 (s_W^2 M_1 + c_W^2 M_2) (\cos2\beta)^2 
   \right], 
\nonumber \\
({\rm m}_{\rm higgs^0}^2)_{23}^\prime &=& ({\rm m}_{\rm higgs^0}^2)_{32}^\prime 
= \frac{ v \cos2\beta}{ \sqrt{2} F } 
    \left[ (m_A^2 - 2 \mu^2)\mu + m_Z^2 (s_W^2 M_1 + c_W^2 M_2) \sin2\beta 
   \right]. 
  \label{eq:HiggsMass1_add}
\end{eqnarray}
%---------------------------<

We define the mass eigenbasis $\phi_i = (\phi_1, \phi_2, \phi_3)$ as 
%--------------------------->
%\begin{eqnarray}
% h_i = S_{i,j} (h_d^0, h_u^0, s_X)_j
%  \label{eq:higgsMassbase}
%\end{eqnarray}
%---------------------------<
%--------------------------->
\begin{eqnarray}
\phi_i &=& S_{ij} h_j,
  \label{eq:rotationHiggs}
\end{eqnarray}
%---------------------------<
where $h_i = (h, H, s_X)$. The mass terms are written as 
%--------------------------->
\begin{eqnarray}
 \mathcal{L} &\supset& - \frac{1}{2} m_i^2 \phi_i^2, 
  \label{eq:HiggsMass}
\end{eqnarray}
%---------------------------<
where $m_{1,2,3}$ are 
%$(m_h, ~ m_H, ~m_X)$ 
in ascending order. 
%When we consider only up to $\mathcal{O}(1/F)$, 
These masses are not different from the diagonal elements of Eq.~\eqref{eq:HiggsMass1} up to $\mathcal{O}(1/F)$, i.e,  $m_h$ and $m_H$ are the same as the light and heavy Higgs boson masses of MSSM, respectively.

This approximation is not valid when $g^2 < (m_{\rm SUSY}^2/F)^2$ as $\mathcal{O}(1/F^2)$ contributions to the lightest Higgs boson mass cannot be negligible. For example, the tree level lightest Higgs boson mass up to $\mathcal{O}(1/F^2)$, in the limit of large $m_X$ (or large $m_A$) and large $\tan \beta$, is obtained as 
\begin{eqnarray}
 m_h^2 &\sim& 
m_Z^2 +  \frac{2 v^2}{F^2} \mu^4 ,
  \label{eq:HiggsMass_F2-ma}
\end{eqnarray}
where we have neglected terms which are proportional to gauge coupling in $\mathcal{O}(1/F^2)$ terms. Thus, if $\mu/\sqrt{F} \sim 0.5$ the lightest Higgs mass can be $\sim 125$ GeV at tree level.

Therefore we include $\mathcal{O}(1/F^2)$ terms only in the neutral Higgs boson mass matrices in our numerical analysis in Section 5. The $\mathcal{O}(1/F^2)$ terms affect the value of the lightest Higgs mass only for large values of $\mu$. If $\mu$ term is very large, the obtained lightest Higgs boson mass is larger than the observed Higgs mass. However, in a general low-scale SUSY breaking scenario additional higher dimensional terms which do not include goldstino superfield can contribute to the Higgs mass. If there are such additional terms, the bound would change.

\subsection{Pseudo scalar mass matrix}

The pseudo scalar mass matrix is written as 
%--------------------------->
\begin{eqnarray}
 \mathcal{L} &\supset& - \frac{1}{2}
\left( 
A_u ~ A_d ~ a_X 
\right)
\left( 
\begin{array}{ccc}
      m_A^2 \cos^2\beta & m_A^2 \sin\beta \cos\beta & \frac{m_A^2 - 2 \mu^2}{\sqrt{2} F}\mu v \cos\beta \\
      m_A^2 \sin\beta \cos\beta & m_A^2 \sin^2\beta & \frac{m_A^2 - 2 \mu^2}{\sqrt{2} F}\mu v \sin\beta \\ 
      \frac{m_A^2 - 2 \mu^2}{\sqrt{2} F}\mu v \cos\beta & \frac{m_A^2 - 2 \mu^2}{\sqrt{2} F}\mu v \sin\beta & m_X^2\\ 
\end{array}
\right)
\left(
\begin{array}{c}
       A_u \\
       A_d \\ 
       a_X   \\ 
\end{array}
\right), ~~~~~
  \label{eq:PseudoMass0}
\end{eqnarray}
%---------------------------<
up to $\mathcal{O}(1/F)$. At this order, would-be Nambu-Goldstone boson is the same as the usual MSSM, $G^0 = \cos\beta A_d - \sin\beta A_u$, then, by the rotation 
%--------------------------->
\begin{eqnarray}
\left(
\begin{array}{c}
        A_d \\
        A_u \\ 
        a_X \\
\end{array}
\right),
&=& 
\left(
\begin{array}{ccc}
       \cos\beta & \sin\beta & 0 \\ 
      -\sin\beta & \cos\beta & 0 \\ 
               0 &         0 & 1 \\ 
\end{array}
\right)
\left(
\begin{array}{c}
        G^0 \\
        A \\ 
        a_X \\
\end{array}
\right),
  \label{eq:rotationPseudo1}
\end{eqnarray}
%---------------------------<
Eq.~\eqref{eq:PseudoMass0} is rewritten as 
%--------------------------->
\begin{eqnarray}
 \mathcal{L} &\supset& - \frac{1}{2}
\left( 
A ~ a_X 
\right)
\left( 
\begin{array}{cc}
      m_A^2 & \frac{m_A^2 - 2 \mu^2}{\sqrt{2} F} \mu v\\
      \frac{m_A^2 - 2 \mu^2}{\sqrt{2} F} \mu v & m_X^2\\ 
\end{array}
\right)
\left(
\begin{array}{c}
       A \\ 
       a_X \\ 
\end{array}
\right). 
  \label{eq:PseudoMass1}
\end{eqnarray}
%---------------------------<

The mass eigenbasis $a_i = (a_1, a_2)$ is defined as 
%--------------------------->
\begin{eqnarray}
a_i &=& A_{ij} A_j,
  \label{eq:rotationPseudoHiggs}
\end{eqnarray}
%---------------------------<
where $A_i = (A, a_X)$. Then, the mass terms are written as 
%--------------------------->
\begin{eqnarray}
 \mathcal{L} &\supset& - \frac{1}{2} m_{ai}^2 a_i^2, 
  \label{eq:HiggsMassA}
\end{eqnarray}
%---------------------------<
where $m_{a 1,2} \,(m_A,~m_{aX}=m_X)$ are in ascending order. 
Thus, the pseudo-scalar Higgs mass is the same as the MSSM pseudo scalar Higgs mass up to $\mathcal{O}(1/F)$. For example, when  $m_A < m_{aX}$, 
%--------------------------->
\begin{eqnarray}
A_{ij}
&=& 
\left(
\begin{array}{cc}
       \cos\theta_a & -\sin\theta_a \\ 
       \sin\theta_a & \cos\theta_a \\ 
\end{array}
\right)_{ij}, \qquad 
 \tan 2\theta_a = - \frac{\sqrt{2} \mu v}{F} \frac{m_A^2 - 2 \mu^2}{m_A^2 - m_X^2}.
  \label{eq:rotationPseudo2}
\end{eqnarray}
%---------------------------<

\subsection{Charged scalar mass matrix}

The charged scalar mass matrix is written as 
%--------------------------->
\begin{eqnarray}
 \mathcal{L} &\supset& - 
\left( 
H_u^+ ~ H_d^+ 
\right)
\left( 
\begin{array}{cc}
      (m_A^2 + m_W^2) \cos^2\beta & (m_A^2 + m_W^2) \sin\beta \cos\beta\\
      (m_A^2 + m_W^2) \sin\beta \cos\beta & (m_A^2 + m_W^2) \sin^2\beta\\ 
\end{array}
\right)
\left(
\begin{array}{c}
       H_u^- \\
       H_d^- \\ 
\end{array}
\right), 
  \label{eq:ChargedMass0}
\end{eqnarray}
%---------------------------<
up to $\mathcal{O}(1/F)$ and this is the same as the charged Higgs mass in MSSM.Eq.~\eqref{eq:ChargedMass0} can be redefined in terms of the would-be Nambu-Goldstone boson $G^-$ and the physical charged Higgs boson $H^-$ by the following rotation 
%--------------------------->
\begin{eqnarray}
\left(
\begin{array}{c}
        H_u^-\\
        H_d^-\\ 
\end{array}
\right)
&=& 
\left(
\begin{array}{cc}
       \cos\beta & - \sin\beta \\ 
       \sin\beta & \cos\beta \\ 
\end{array}
\right)
\left(
\begin{array}{c}
        H^-\\
        G^-\\ 
\end{array}
\right),
  \label{eq:MSSMCharged}
\end{eqnarray}
%---------------------------<
yielding the mass term 
%--------------------------->
\begin{eqnarray}
 \mathcal{L} &\supset& - (m_A^2 + m_W^2) H^+H^-, 
  \label{eq:ChargedMass}
\end{eqnarray}
%---------------------------<
up to $\mathcal{O}(1/F)$.\footnote{
However, if we take into account higher orders in $1/F$ expansion, the mixing angle would change. 
%in the charged Higgs boson sector as well.
}
%

%%%%%%%%%%%%%%%%%%%%%%%%%%%%%%%%%%%%%%%%%%%%%%%%%%%%%%%%%%%%%%%%%%%%%%%%%%%% 
%  gaugino Higgsino goldstino   
%%%%%%%%%%%%%%%%%%%%%%%%%%%%%%%%%%%%%%%%%%%%%%%%%%%%%%%%%%%%%%%%%%%%%%%%%%%% 
\section{Gaugino-Higgsino-Goldstino mass matrices}

Through the electroweak symmetry breaking, the fermionic component of the goldstino superfield mixes with gauginos and Higgsinos. In this section, we write the neutralino and chargino mass matrices and define their mass eigenstates.

\subsection{Neutralino mass matrix}

From Eq.~\eqref{eq:Lagrangian0}, the neutralino mass terms are obtained as 
%--------------------------->
\begin{eqnarray}
 \mathcal{L} &\supset& - \frac{1}{2} 
\left( 
\tilde{B} ~ \tilde{W} ~ \tilde{H}_d^0 ~ \tilde{H}_u^0 ~ \psi_X
\right)
M_{\tilde{N}}
\left(
\begin{array}{c}
       \tilde{B} \\
       \tilde{W} \\ 
   \tilde{H}_d^0 \\ 
   \tilde{H}_u^0 \\ 
         \psi_X  \\ 
\end{array}
\right) + {\rm h.c. }, 
  \label{eq:NeutralinoMass0}
\\
M_{\tilde{N}} &=&
\left( 
\begin{array}{ccccc}
      M_1 & 0   & -m_Z s_W \cos\beta &  m_Z s_W \sin\beta & (M_{\tilde{N}})_{15} \\
      0   & M_2 &  m_Z c_W \cos\beta & -m_Z c_W \sin\beta & (M_{\tilde{N}})_{25} \\ 
      -m_Z s_W \cos\beta &  m_Z c_W \cos\beta & 0   & \mu & (M_{\tilde{N}})_{35} \\ 
       m_Z s_W \sin\beta & -m_Z c_W \sin\beta & \mu & 0   & (M_{\tilde{N}})_{45} \\ 
      (M_{\tilde{N}})_{51} & (M_{\tilde{N}})_{52} & (M_{\tilde{N}})_{53} & (M_{\tilde{N}})_{54} & 0 \\ 
\end{array}
\right),
\nonumber 
\end{eqnarray}
%---------------------------<
where 
%--------------------------->
\begin{eqnarray}
(M_{\tilde{N}})_{15} &=& (M_{\tilde{N}})_{51} = \frac{1}{8 \sqrt{2} F} v \cos2\beta s_W m_Z M_1, 
\nonumber \\ 
(M_{\tilde{N}})_{25} &=& (M_{\tilde{N}})_{52} =-\frac{1}{8 \sqrt{2} F} v \cos2\beta c_W m_Z M_2, 
\nonumber \\ 
(M_{\tilde{N}})_{35} &=& (M_{\tilde{N}})_{53} = 
 \frac{1}{ 2 \sqrt{2} F} v \cos\beta \left( \mu^2 + \frac{1}{2} m_Z^2 \cos2\beta \right), 
\nonumber \\ 
(M_{\tilde{N}})_{45} &=& (M_{\tilde{N}})_{54} = 
 \frac{1}{ 2 \sqrt{2} F} v \sin\beta \left( \mu^2 - \frac{1}{2} m_Z^2 \cos2\beta \right), 
  \label{eq:NeutralinoMass1}
\end{eqnarray}
%---------------------------<
up to $\mathcal{O}(1/F)$. 
We write the mass eigenbasis as $\tilde{\chi} = (\tilde{\chi}_0,\tilde{\chi}_1,\tilde{\chi}_2,\tilde{\chi}_3,\tilde{\chi}_4)^T$ where $m_{\tilde{\chi}_i} < m_{\tilde{\chi}_j}$ with $i<j$ and $\tilde{\chi}_0$ corresponds to goldstino. It is defined as 
%--------------------------->
\begin{eqnarray}
\tilde{\chi}_i = N_{ij} \tilde{N}^0_j = \xi_i N^\prime_{ij} \tilde{N}^0_j, 
  \label{eq:NeutralinoMass1}
\end{eqnarray}
%---------------------------<
where $N^\prime_{ij}$ is a rotation matrix which diagonalizes the mass matrix and $\tilde{N}^0 = (\tilde{B}, \tilde{W}, \tilde{H}_d^0, \tilde{H}_u^0, \psi_X)^T$, respectively. The $\xi_i$ is $1$ ($i$) for positive (negative) eigenvalues of the diagonalized mass matrix.  The mass eigenvalues are the same as MSSM with massless goldstino up to $\mathcal{O}(1/F)$.

\subsection{Chargino mass matrix}

The chargino mass matrix is the same as that of MSSM up to $\mathcal{O}(1/F)$:
%--------------------------->
\begin{eqnarray}
 \mathcal{L} &\supset& -  
\left( \tilde{W}^+ ~ \tilde{H}_u^+ 
\right)
\left( 
\begin{array}{cc}
      M_2 & \sqrt{2} m_W \cos\beta \\
      \sqrt{2} m_W \sin\beta   & -\mu \\ 
\end{array}
\right)
\left(
\begin{array}{c}
       \tilde{W}^- \\ 
     \tilde{H}_d^- \\ 
\end{array}
\right) + {\rm h.c. }. 
  \label{eq:CharginoMass0}
\end{eqnarray}
%---------------------------<
We describe the mass eigenstates as $\tilde{\chi}^- = (\tilde{\chi}^-_1, \tilde{\chi}^-_2)^T$ where $m_{\tilde{\chi}^-_1} < m_{\tilde{\chi}^-_2}$ and defined as 
%--------------------------->
\begin{eqnarray}
\tilde{\chi}^-_{L i} = C^L_{ij} \tilde{C}^-_{L j} 
\quad {\rm and} \quad 
\tilde{\chi}^-_{R i} = C^R_{ij} \tilde{C}^-_{R j} 
                     = \epsilon_i C^{R \prime}_{ij} \tilde{C}^-_{R j}.
  \label{eq:CharginoMass1}
\end{eqnarray}
%---------------------------<
$C^L_{ij}$ and $C^{R \prime}_{ij}$ are the rotation matricies which diagonalize the mass matrix and $\tilde{C}^-_{L} = (\tilde{W}^-, \tilde{H}_d^-)_L^T$ and 
$\tilde{C}^-_{R} = (\tilde{W}^-, \tilde{H}_u^-)_R^T$, respectively. $\epsilon_i$ is $1$ ($-1$) for positive (negative) eigenvalues of the diagonalized mass matrix obtained by using  $C^L_{ij}$ and $C^{R}_{ij}$.

%%%%%%%%%%%%%%%%%%%%%%%%%%%%%%%%%%%%%%%%%%%%%%%%%%%%%%%%%%%%%%%%%%%%%%%%%%%% 
%  Production & Decay width  
%%%%%%%%%%%%%%%%%%%%%%%%%%%%%%%%%%%%%%%%%%%%%%%%%%%%%%%%%%%%%%%%%%%%%%%%%%%% 
\section{Production and Decay of the sgoldstino}

We now turn to study the production and decay of sgoldstino at the LHC. First, we discuss the partial widths of sgoldstino and pseudo-sgoldstino using approximations. Then, we present the numerical results for production cross section and branching ratios.

\subsection{Partial decay widths}

In this subsection, we discuss the partial decay widths of sgoldstino and pseudo-sgoldstino assuming these are much heavier than Z boson and mixing with MSSM Higgs bosons is not large, for simplicity.  The full analytical expressions for the partial widths are compiled in Appendix C.

\subsubsection*{Gauge boson branch}

The partial decay width to a pair of gluons $gg$ which contributes not only to the decay but also to the production at the LHC is obtained to be 
%--------------------------->
\begin{eqnarray}
 \Gamma(\phi \to gg) &\approx& \frac{1}{4 \pi} \frac{M_3^2 m_{\phi}^3}{F^2}, 
  \label{eq:Width_gg_app}
\end{eqnarray}
%---------------------------<
where 
%MA
$\phi=s,a$ 
\footnote{The $s$($a$) denotes a sgoldstino(pseudo-sgoldstino)-dominant particle in $\phi_i$($a_i$), which are defined in Eq.~\eqref{eq:rotationHiggs}(Eq.~\eqref{eq:rotationPseudoHiggs}).}.
Then we can obtain the following relation,
%
%--------------------------->
\begin{eqnarray}
 && \Gamma(\phi \to gg) : \Gamma(\phi \to \gamma \gamma) : \Gamma(\phi \to \gamma Z) 
\nonumber \\
&\approx& M_3^2 : \frac{1}{8} \left( c_W^2 M_1 + s_W^2 M_2 \right)^2 : \frac{1}{4} s_W^2 c_W^2 \left( M_1 - M_2 \right)^2. 
  \label{eq:Width_2photon_app}
\end{eqnarray}
%---------------------------<
For massive boson final states, if the transverse modes dominate, the partial decay widths is obtained to be 
%--------------------------->
\begin{eqnarray}
 && \Gamma(\phi \to gg) : \Gamma(\phi \to W_T W_T) : \Gamma(\phi \to Z_T Z_T) 
\nonumber \\
&\approx& M_3^2 : \frac{1}{4} M_2^2 : \frac{1}{8} \left( s_W^2 M_1 + c_W^2 M_2 \right)^2. 
  \label{eq:Width_WWT_app}
\end{eqnarray}
%---------------------------<
On the other hand, if the longitudinal mode is dominant, the partial decay widths can be obtained by the would-be Goldstone boson interaction through the equivalence theorem. The interactions of sgoldstino with the would-be Goldstone boson $G^0$ is given by 
%--------------------------->
\begin{eqnarray}
 \mathcal{L} &\supset& 
\frac{1}{2 \sqrt{2} F} \left[ 2 \mu^3 \sin2\beta 
                             +m_Z^2 \left( s_W^2 M_1 + c_W^2 M_2 \right) (\cos2\beta)^2
                      \right] s_X G^0 G^0,  
  \label{eq:L_sGG}
\end{eqnarray}
%---------------------------<
up to $\mathcal{O}(1/F)$. After dropping the term proportional to $m_Z^2$, the decay width is obtained to be 
%--------------------------->
\begin{eqnarray}
\Gamma(s \to G^0 G^0) &\approx& 
\frac{1}{8 \pi m_{s}} \left[ \frac{1}{2 \sqrt{2} F} ( 2 \mu^3 \sin2\beta ) \right]^2
\nonumber \\
 &=&  \frac{1}{16 \pi} \frac{\mu^6}{m_s F^2} (\sin2\beta)^2. 
  \label{eq:Width_G0G0_app}
\end{eqnarray}
%---------------------------<
The ratio of partial decay widths
%--------------------------->
\begin{eqnarray}
\Gamma(s_X \to G^0 G^0): \Gamma(s_X \to W_L W_L) : \Gamma(s_X \to Z_L Z_L) \approx 1:2:1.
  \label{eq:Width_WWL_app}
\end{eqnarray}
%---------------------------<
There is  no pseudo-sgoldstino interactions with $G^0G^0$ in the absence of CP violation.

%%%%%%
\subsubsection*{Higgs boson branch}

Assuming $m_Z \ll (\mu$ and $m_A)$ and $m_Z M_a \ll \mu^2$, the decay width of sgoldstino to a pair of  lightest $CP$-even higgs $h$ is
%
%--------------------------->
\begin{eqnarray}
\Gamma(s \to hh) &\approx& \frac{1}{8 \pi m_{s}} \left[ \frac{\mu }{2 \sqrt{2}F} \left\{ 
           (m_A^2 - 2 \mu^2) \sin2\alpha + m_A^2 \sin2\beta \right\} \right]^2
\nonumber \\
 &\approx&  \frac{1}{16 \pi} \frac{\mu^6}{m_{s} F^2} (\sin2\beta)^2.
  \label{eq:Width_hh_app}
\end{eqnarray}
%---------------------------<
The second line of Eq.~\eqref{eq:Width_hh_app} can be obtained by using $(\sin2\alpha) \sim -(\sin2\beta)$. In such a limit, the interactions $s_X hh$ and $s_X G^0 G^0$ are the same at the leading order. Then, the following relation is obtained 
%--------------------------->
\begin{eqnarray}
\Gamma(s \to hh): \Gamma(s \to W_L W_L) : \Gamma(s \to Z_L Z_L) \approx 1:2:1. 
  \label{eq:Width_hh_WWL_app}
\end{eqnarray}
%---------------------------<
On the other hand, the pseudo-sgoldstino does not decay into $hh$ in the absence of CP violation.

If kinematically allowed, decays to other Higgs bosons also exist. By the same approximation used to derive Eq.~\eqref{eq:Width_hh_app}, the decay widths of $s_X$ to heavy Higgs bosons are
% 
%--------------------------->
\begin{eqnarray}
\Gamma(s \to HH) \approx \frac{1}{16 \pi} \frac{\mu^2(m_A^2-\mu^2)^2}{m_{s} F^2} (\sin2\beta)^2, 
  \label{eq:Width_phiphi_app}
\end{eqnarray}
%---------------------------<
and 
\begin{eqnarray}
\Gamma(s \to H^+H^-)/2 \approx \Gamma(s \to AA) \approx \Gamma(s \to HH), 
  \label{eq:Width_phiphi_app-comp}
\end{eqnarray}
%---------------------------<
where we have assumed $m_s \gg m_A$ for simplicity. 
On the other hand, 
%--------------------------->
\begin{eqnarray}
\Gamma(s \to hH) &\approx& \frac{1}{32 \pi} \frac{\mu^2(m_A^2-2\mu^2)^2}{m_s F^2} (\cos2\beta)^2, 
\nonumber \\
\Gamma(a \to hA) &\approx& \frac{1}{32 \pi} \frac{\mu^2(m_A^2-2\mu^2)^2}{m_a F^2}.
  \label{eq:Width_hH_hA_app}
\end{eqnarray}
%---------------------------<
Note that there is no $1/\tan\beta$ suppression in Eq.~\eqref{eq:Width_hH_hA_app}. Thus, the partial width of $s_X \to hH$  is larger than the other Higgs boson branches and the longitudinal mode of $WW/ZZ$ in the limit of large $\tan\beta$.　

%%%%%%
\subsubsection*{Fermion and sfermion branch}

Sgoldstino interactions with SM fermions is proportional to $ m_f A_f / (y_f F)$ as shown in Eq.~\eqref{eq:L_eff_5_MSSM} in Appendix A. However, sgoldstino-fermion-fermion couplings originating from mixing with MSSM Higgs bosons can contribute at the same order. In the limit $m_Z \ll m_A \ll m_X$ or $m_Z \ll m_X \ll m_A$, $(\sin2\alpha) \sim (-\sin2\beta)$, the decay widths of sgoldstino to SM fermions take the form 
%--------------------------->
\begin{eqnarray}
 \Gamma(s \to \bar{t}t) &\approx& \frac{3}{16\pi} \frac{m_s m_t^2}{F^2} 
       \left[ \frac{A_t}{y_t} + 2 \frac{\mu^3}{m_s^2} \sin2\beta 
              -\left( \frac{m_A^2-2\mu^2}{m_s^2-m_A^2} \right)
                \frac{\mu \cos2\beta}{\tan\beta}
      \right]^2, 
\nonumber \\ 
 \Gamma(s \to \bar{b}b) &\approx& \frac{3}{16\pi} \frac{m_s m_b^2}{F^2} 
       \left[ \frac{A_b}{y_b} + 2 \frac{\mu^3}{m_s^2} \sin2\beta 
              +\left( \frac{m_A^2-2\mu^2}{m_s^2-m_A^2} \right) 
                \mu \cos2\beta \tan\beta
      \right]^2. 
  \label{eq:Width_sx_fermion_app}
\end{eqnarray}
Note that the third term in the expression for $\Gamma(s_X \to \bar{b}b)$ in Eq.~\eqref{eq:Width_sx_fermion_app} are $\tan\beta$ enhanced.
In the same limit as above, the decay widths of pseudo-sgoldstino to SM fermions is written as 
%--------------------------->
\begin{eqnarray}
 \Gamma(a \to \bar{t}t) &\approx& \frac{3}{16\pi} \frac{m_a m_t^2}{F^2} 
       \left[ \frac{A_t}{y_t} 
             +\left( \frac{m_A^2-2\mu^2}{m_a^2-m_A^2} \right) \frac{\mu}{\tan\beta}
      \right]^2, 
\nonumber \\ 
 \Gamma(a \to \bar{b}b) &\approx& \frac{3}{16\pi} \frac{m_a m_b^2}{F^2} 
       \left[ \frac{A_b}{y_b} 
             +\left( \frac{m_A^2-2\mu^2}{m_a^2-m_A^2} \right) \mu \tan\beta 
      \right]^2. 
  \label{eq:Width_ax_fermion_app} 
\end{eqnarray}
%---------------------------<
Similar to the case of $\Gamma(s \to \bar{b}b)$, there is $\tan\beta$ enhancement arising from mixing in $\Gamma(a \to \bar{b}b)$. Estimating the width of the tau branch is straightforward.

Next, we discuss partial widths for sfermion final states. As shown in Appendix A, $s_X \tilde{f}_L \tilde{f}_L$ and $s_X \tilde{f}_R \tilde{f}_R$ couplings are proportional $v^2/F$. On the other hand, the $\phi \tilde{f}_L \tilde{f}_R$ couplings are proportional $v$, thus making them larger than $s_X \tilde{f}_L \tilde{f}_L$ and $s_X \tilde{f}_R \tilde{f}_R$ couplings. Assuming left-right mixing is small in the sfermion sector, 
%--------------------------->
\begin{eqnarray}
 \Gamma(s \to \tilde{t}_1^* \tilde{t}_2 + \tilde{t}_1 \tilde{t}_2^* ) &\approx& 
     \frac{3}{16\pi} \frac{m_t^2}{m_s F^2} \frac{1}{\tan^2\beta} \left[  
       \left(\frac{ A_t}{y_t} \mu + \frac{1}{2} m_A^2 \sin2\beta \right)  
     - 2 \frac{\mu^3}{m_s^2} \sin2\beta\left(\frac{A_t}{y_t}\tan\beta +\mu \right) 
\right . \nonumber \\ 
&& \left . \qquad \qquad \qquad \qquad 
+  \left(\frac{m_A^2-2\mu^2}{m_s^2-m_A^2}\mu\cos2\beta\right) \left( \frac{A_t}{y_t} - \mu \tan\beta \right)
  \right]^2,
  \label{eq:Width_stop_app1}
\end{eqnarray}
%---------------------------<
where kinetic suppression is neglected assuming $m_{\tilde{t}_{1(2)}} \ll m_s$. On the other hand, if the mixing is maximally large, 
%--------------------------->
\begin{eqnarray}
 \Gamma(s \to \tilde{t}_1^* \tilde{t}_1) &\approx&  \frac{3}{32\pi} \frac{m_t^2}{m_s F^2} \frac{1}{\tan^2\beta} \left[
     \left( \frac{A_t}{y_t} \mu + \frac{1}{2} m_A^2 \sin2\beta \right)  
    - 2 \frac{\mu^3}{m_s^2} \sin2\beta\left(\frac{A_t}{y_t}\tan\beta +\mu \right) 
\right . \nonumber \\ 
&& \left . \qquad \qquad \qquad \qquad 
+  \left(\frac{m_A^2-2\mu^2}{m_s^2-m_A^2}\mu\cos2\beta\right) \left( \frac{A_t}{y_t} - \mu\tan\beta \right)
  \right]^2.
  \label{eq:Width_stop_app2}
\end{eqnarray}
%---------------------------<

In the same limit as above, the partial decay widths of pseudo-sgoldstino to sfermions is given by, 
\begin{eqnarray}
 \Gamma(a \to \tilde{t}_1^* \tilde{t}_2 + \tilde{t}_1 \tilde{t}_2^* ) &\approx& 
     \frac{3}{16\pi} \frac{m_t^2}{m_a F^2} \frac{1}{\tan^2\beta} \left[  
  \left(\frac{ A_t}{y_t} \mu + \frac{1}{2} m_A^2 \sin2\beta \right)  \right . \nonumber \\ 
&& \left . \qquad \qquad \qquad \qquad 
-  \left(\frac{m_A^2-2\mu^2}{m_a^2-m_A^2}\mu\right) \left( \frac{A_t}{y_t} - \mu \tan\beta \right)
  \right]^2,
  \label{eq:Width_psuedo-stop_app1}
\end{eqnarray}

Estimating sbottom and stau branch is straightforward. One of the main difference is $m_t/\tan\beta \rightarrow  m_{b(\tau)} \tan\beta$.

%%%%%%
\subsubsection*{Gaugino-Higgsino-Gravitino branch}

The partial decay width of the gravitino final state can be written as 
%--------------------------->
\begin{eqnarray}
 \Gamma(\phi \to \tilde{G} \tilde{G}) &\approx& \frac{m_{\phi}^5}{32 \pi F^2}, 
  \label{eq:Width_top_app}
\end{eqnarray}
%---------------------------<
which implies that the branching ratio can be large when sgoldstino is heavy. 

Assuming sgoldstino-Higgs mixing is small, we also present the decay width of sgoldstino to pure higgsino final states 
%--------------------------->
\begin{eqnarray}
 \Gamma(\phi \to \tilde{H}^0_1 \tilde{H}^0_2) 
&\approx& 
 \Gamma(\phi \to \tilde{H}^+ \tilde{H}^-) 
 \approx 
\frac{1}{64 \pi} m_{\phi} \frac{m_A^4\sin^2 2\beta}{F^2}, 
  \label{eq:Width_susy1}
\end{eqnarray}
%---------------------------<
where kinetic suppression is neglected assuming sgoldstino is much heavier than higgsino.

%%%%%%%%%%%%%%%%%
%  Production 
%%%%%%%%%%%%%%%%%
\subsection{Production cross section}

Sgoldstino and pseudo-sgoldstino are mainly produced through the gluon fusion process at the LHC. The corresponding decay width is obtained to be $\Gamma(s \to gg) \sim (M_3/F)^2 m_s^3/(4\pi)$, if sgoldstino-MSSM Higgs mixing is not very large. Then, the production cross section of sgoldstino depends on the ratio of gluino mass and $F$, $1/\Lambda = M_3/F$. 

\begin{figure}[!h]
 \begin{center}
  \includegraphics[width=7.5cm]{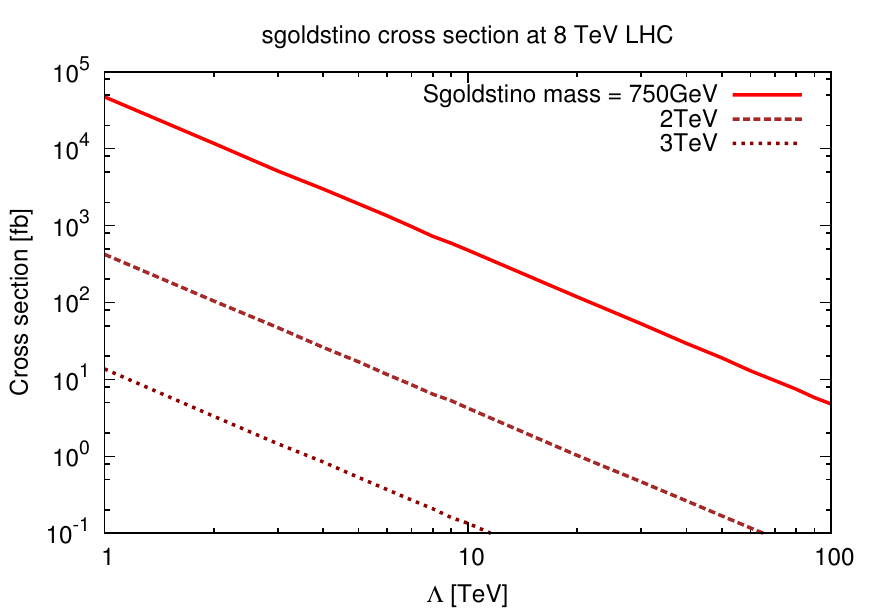}
  \includegraphics[width=7.5cm]{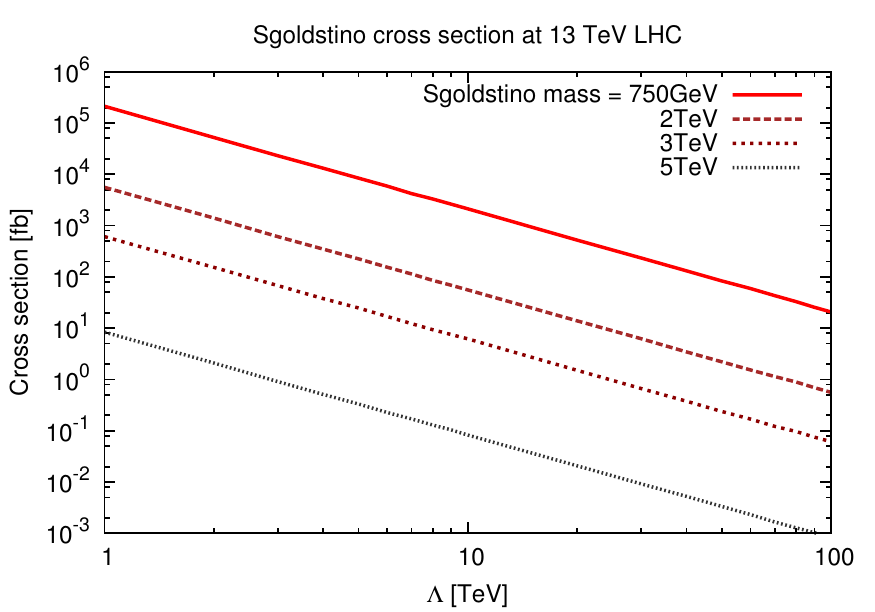}
 \end{center}
\caption{ Production cross section of the sgoldstino $s_X$ at $\sqrt{s}=8$ TeV (left) and $\sqrt{s}=13$ TeV (right). The x-axis, $\Lambda$, denotes $1/\Lambda=M_3/F$. In the left (right) figure, the sgoldstino mass is $3$, $2$ and $0.75$ TeV ($5$, $3$, $2$ and $0.75$ TeV) from below.}  
\label{fig:crosssection_sx}
\end{figure}

The production cross section of sgoldstino is presented in Fig.~\ref{fig:crosssection_sx}. 
To calculate the cross sections we use MadGraph 5 \cite{Alwall:2011uj,Alwall:2014hca} with leading order NNPDF2.3 \cite{Ball:2012cx} and Feynrules \cite{Alloul:2013bka} by approximating the total decay width of sgoldstino to be $\Gamma(s \to gg)$. The case of pseudo-sgoldstino is similar.

%===========================>
%===========================<

%%%%%%%%%%%%%%%%%
%  Branching ratio  
%%%%%%%%%%%%%%%%%
\subsection{Branching ratio}

In the final part of the section we discuss the branching ratios of sgoldstino and pseudo-sgoldstino to various final states. Branching ratios are mostly determined by the ratio of soft masses and $\sqrt{F}$ \footnote{ 
An exceptional example would be the branching to fermion final states, which depends on $v/\sqrt{F}$ as discussed previously. 
}.
The discussion is illustrated using sample points shown in Table \ref{tb:bench} with $\sqrt{F} = 5$ TeV and $\tan\beta =10$. 
Sfermion soft masses are taken to be universal. The A term ($A_f/y_f$) are also taken to be universal which are determined by the requirement of a light Higgs of mass $125$ GeV at each parameter point. 

\begin{table}[h]
\centering
  \begin{tabular}{ l | l | l|l|l|l| }
\cline{2-6}
 & \multicolumn{5}{ c |}{Sample point}\\ \cline{2-6}
\hline
Parameter (in TeV) &  I &   II & III& IV& V \\ \hline
\hline
    $\mu$ & -2 & -2 &-2&-0.2&-2 \\ 
    $m_A$ & 4 & 4 &4&4&0.3 \\ 
    $m_{\tilde{f}}$ & 1 & 1 &2&2&2 \\ 
    $M_3$ & 2 & 2 &2&2&2 \\ 
    $M_2$ & 2 & 0.6 &2&2&2 \\ 
    $M_1$ & 1.5 & 0.3 &1.5&1.5&1.5 \\ \hline
    \hline

  \end{tabular} 
\caption{Sample points.}
\label{tb:bench}
\end{table}

\begin{figure}[!h]
 \begin{center}
  \includegraphics[width=7.5cm]{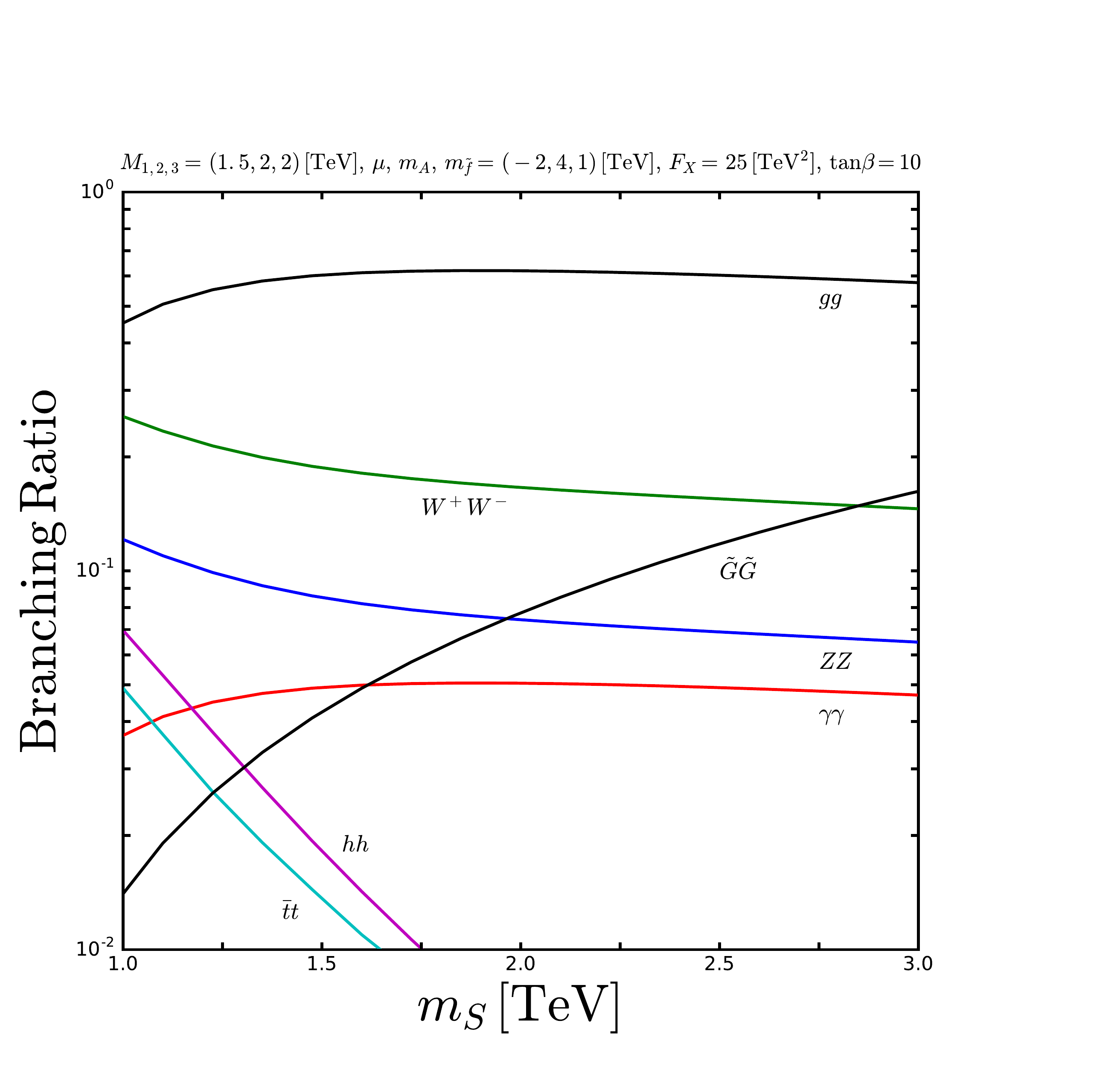}
  \includegraphics[width=7.5cm]{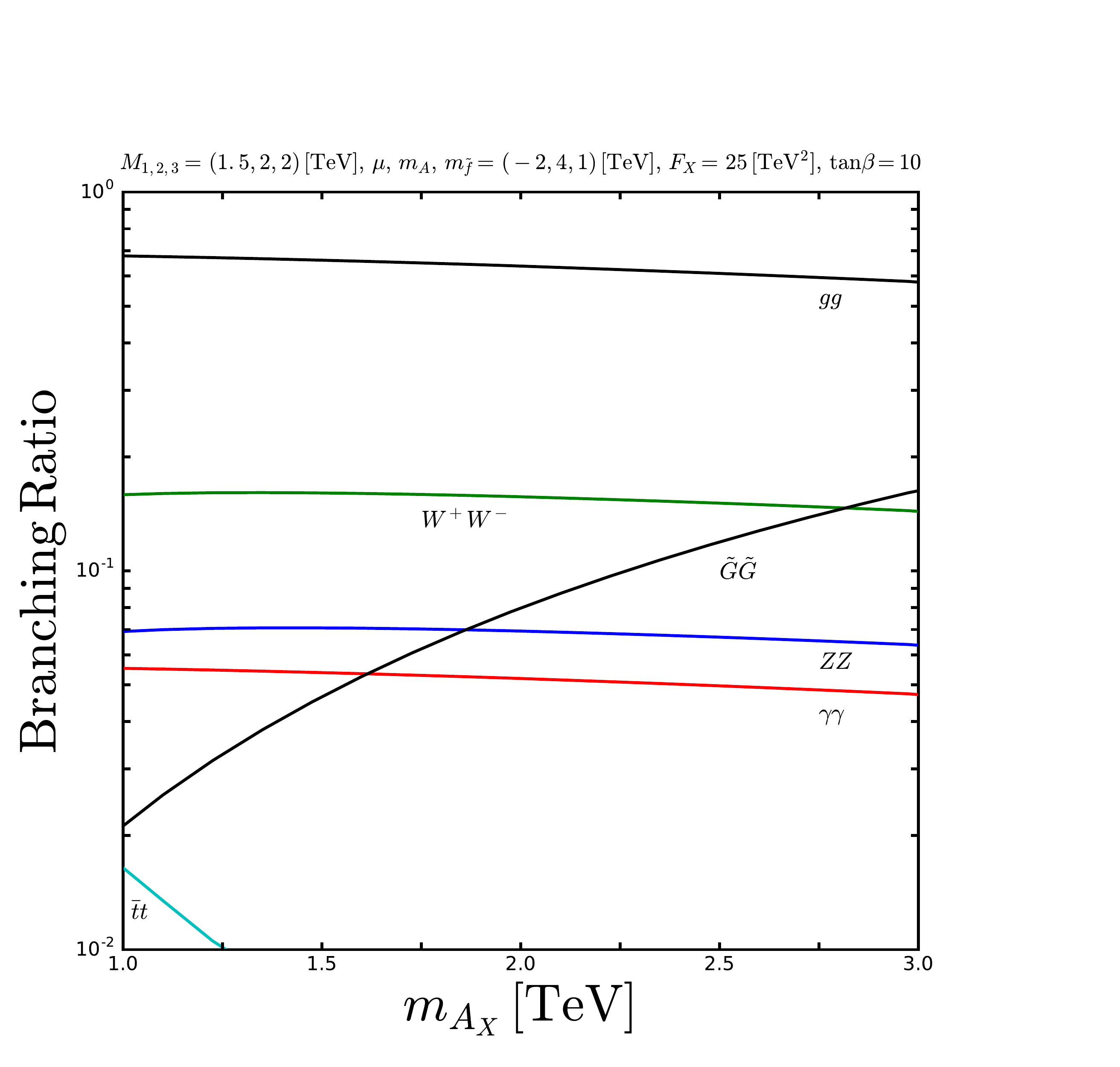}
 \end{center}
\caption{Sample Point I: branching ratio of the sgoldstino (left panel) and pseudo-sgoldstino (right panel) at $\sqrt{F} = 5$ TeV, $(\mu, m_A, m_{\tilde{f}})=(-2,4,1)$ TeV and $(M_3,M_2,M_1) = (2, 2, 1.5)$ TeV with $\tan\beta=10$. 
}  
\label{fig:Br1}
\end{figure}

%%%%%%%%%%%%%%%%%
For sample point I, branching ratios of sgoldstino and pseudo-sgoldstino to various final states are shown in Fig.~\ref{fig:Br1}.
As $\Gamma(\phi \to gg) \propto M_3^2 m_\phi^3/F^2$ and $\Gamma(\phi \to \tilde{G} \tilde{G}) \propto m_\phi^5/F^2$ ( see discussion in Sec 5.1), the branching ratio $\phi \to \tilde{G} \tilde{G}$ becomes large in the heavy sgoldstino (pseudo-sgoldstino) region. 
For small sgoldstino masses, Higgs-sgoldstino mixing becomes prominent (since $\mu$ is large we cannot neglect higgs-sgoldstino mixing) and enhances not only the $hh$ mode but also the longitudinal modes of weak gauge bosons as given by Eqs.~\eqref{eq:Width_hh_app} and \eqref{eq:Width_hh_WWL_app}. 
On the other hand, there is no such enhancements in the case of pseudo-sgoldstino due to the absence of CP violation.

\begin{figure}[!h]
 \begin{center}
  \includegraphics[width=7.5cm]{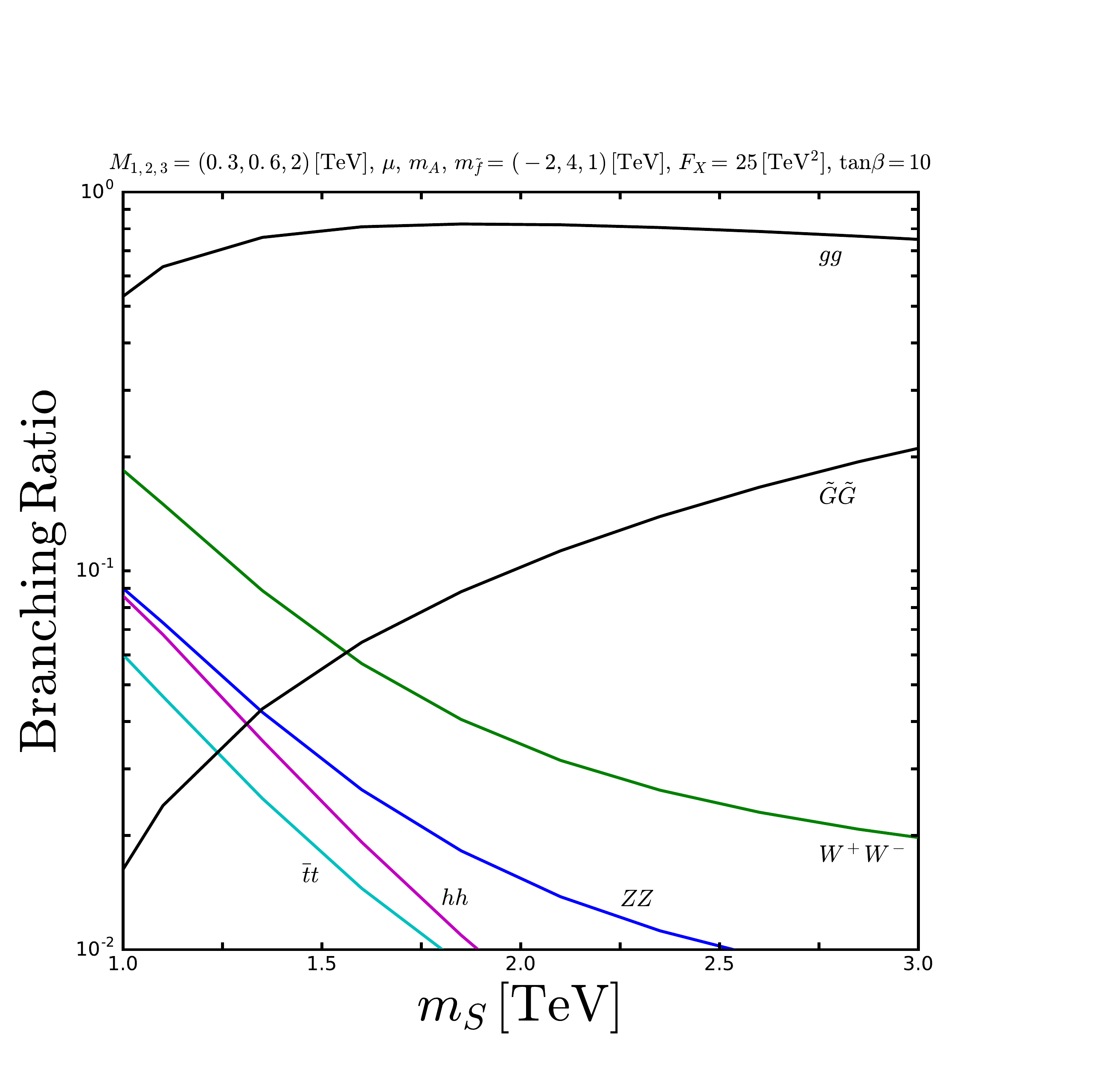}
  \includegraphics[width=7.5cm]{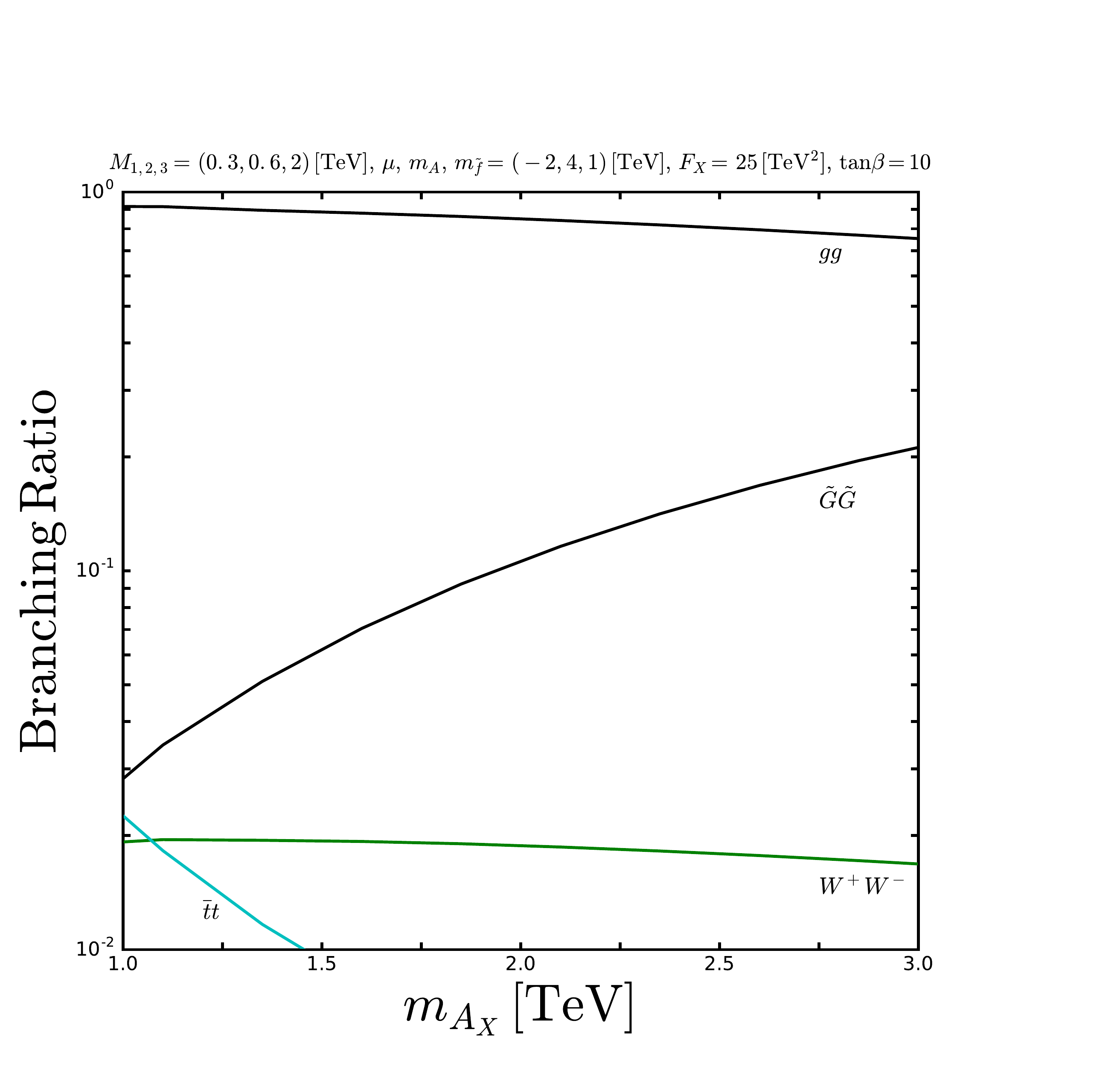}
 \end{center}
\caption{ Sample Point II: branching ratio of the sgoldstino (left panel) and pseudo-sgoldstino (right panel) at $\sqrt{F} = 5$ TeV, $(\mu, m_A, m_{\tilde{f}})=(-2,4,1)$ TeV and $(M_3,M_2,M_1) = (2, 0.6, 0.3)$ TeV with $\tan\beta=10$. 
}  
\label{fig:Br2}
\end{figure}

Since the partial widths for transverse gauge boson modes can be written in the form of Eqs.~\eqref{eq:Width_2photon_app} and \eqref{eq:Width_WWT_app}, it is easy to understand how the branching ratios change with the variation of gaugino masses. Sample point II differs from sample point I only with respect to gaugino masses, where $M_2$ is 0.3 times $M_2$ of sample point I and  $M_1$ is 0.2 times $M_1$ of sample point I, respectively.
The results are presented in Fig.~\ref{fig:Br2} for sgoldstino and pseudo-sgoldstino.

%===========================>
%===========================<
%===========================>
%===========================<

%%%%%%%%%%%%%%%%%

\begin{figure}[!h]
 \begin{center}
  \includegraphics[width=7.5cm]{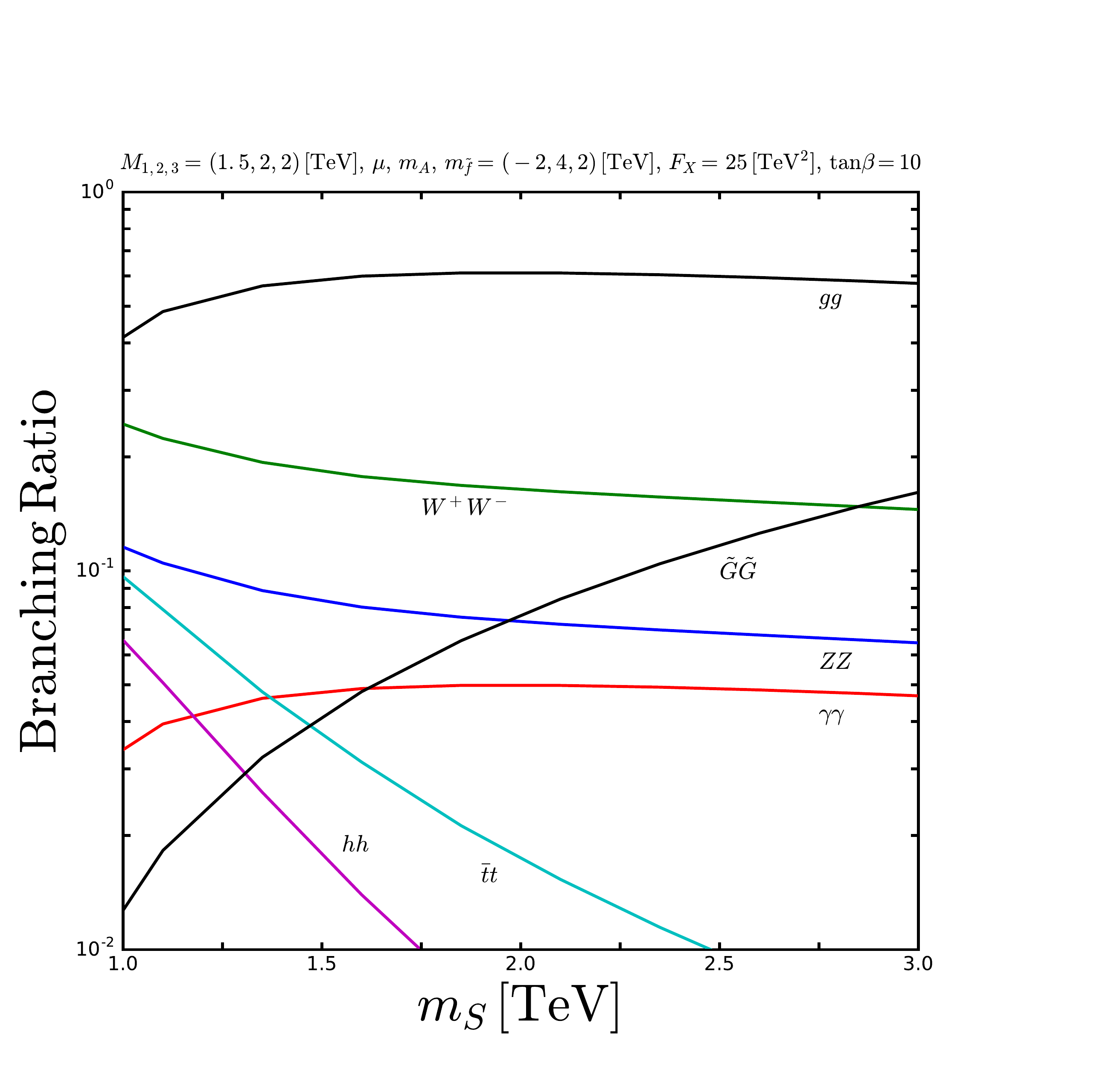}
  \includegraphics[width=7.5cm]{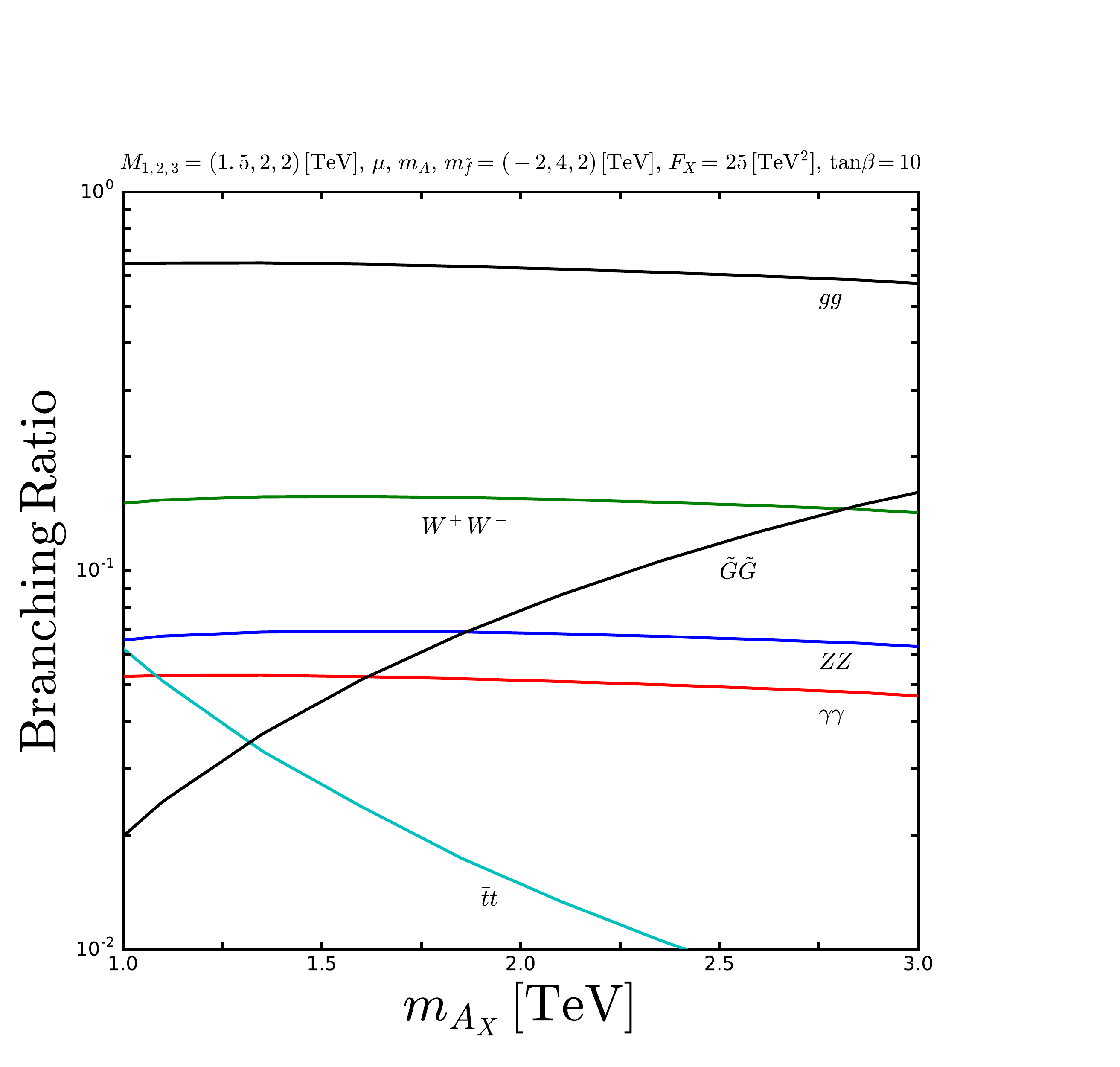}
 \end{center}
\caption{ Sample Point III: branching ratio of the sgoldstino (left panel) and pseudo-sgoldstino (right panel) at $\sqrt{F} = 5$ TeV, $(\mu, m_A, m_{\tilde{f}})=(-2,4,2)$ TeV and $(M_3,M_2,M_1) = (2, 2, 1.5)$ TeV with $\tan\beta=10$. 
}  
\label{fig:Br_heavystop}
\end{figure}

In Fig.~\ref{fig:Br_heavystop}, we show the branching ratio of sgoldstino and pseudo-sgoldstino for sample point III. Here, sfermion masses ($m_{\tilde{f}}$) are set to $2$ TeV instead of $1$ TeV in sample point I. This change impacts the ratio $A/\sqrt{F}$ by making it large thereby enhancing the $t\bar{t}$ mode which depends on $ m_f A_f / (y_f F)$ as prescribed in Eq. \eqref{eq:Width_sx_fermion_app} and Eq. \eqref{eq:Width_ax_fermion_app}. 

%
%===========================>
%===========================<

\begin{figure}[!h]
 \begin{center}
  \includegraphics[width=7.6cm]{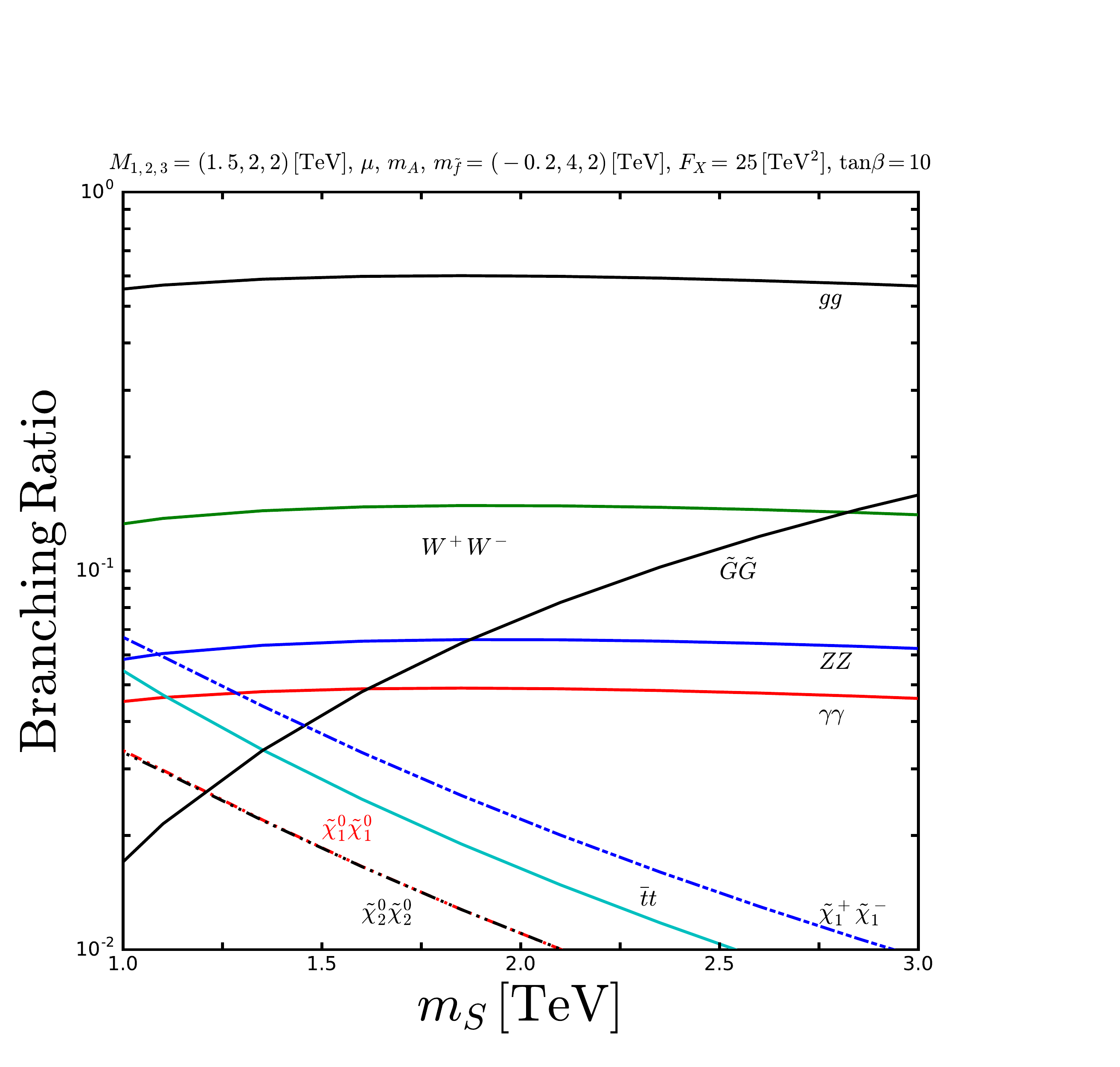}
  \includegraphics[width=7.6cm]{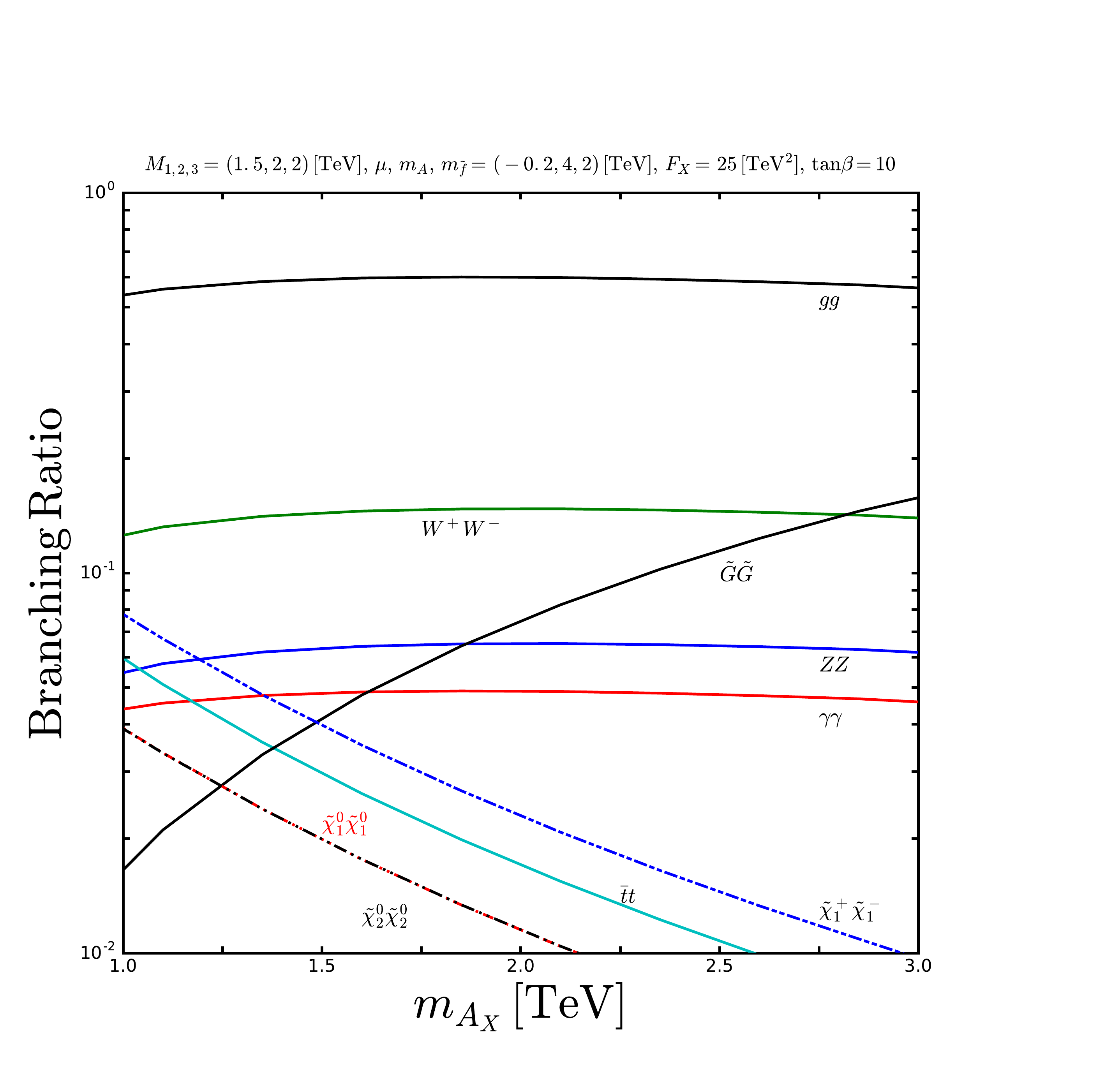}
 \end{center}
\caption{Sample Point IV: branching ratio of the sgoldstino (left panel) and pseudo-sgoldstino (right panel) at $\sqrt{F} = 5$ TeV, $(\mu, m_A, m_{\tilde{f}})=(-0.2,4,2)$ TeV and $(M_3,M_2,M_1) = (2, 2, 1.5)$ TeV with $\tan\beta=10$. 
}  
\label{fig:Br_smallmu}
\end{figure}

%%%%%%%%%%%%%%%%%
We also consider the case of small $\mu$ (sample point IV), where $|\mu|$ is $0.2$ TeV instead of $2$ TeV as in sample point I. The results are depicted in Fig.~\ref{fig:Br_smallmu}.
Since Higgs-sgoldstino mixing depends on the value of the $\mu$, 
the branching ratio of $hh$ and longitudinal modes of $WW$/$ZZ$ is not large, see Eqs.~\eqref{eq:Width_hh_app} and \eqref{eq:Width_hh_WWL_app}. On the other hand, small values of $\mu$ results in light higgsino masses, thus this channel is kinematically allowed. Branching to Higgsino final states can be large since the decay width depends on $m_A^4/F^2$ as shown in Eq. \eqref{eq:Width_susy1}.

%
%===========================>
%===========================<

\begin{figure}[!h]
 \begin{center}
  \includegraphics[width=7.5cm]{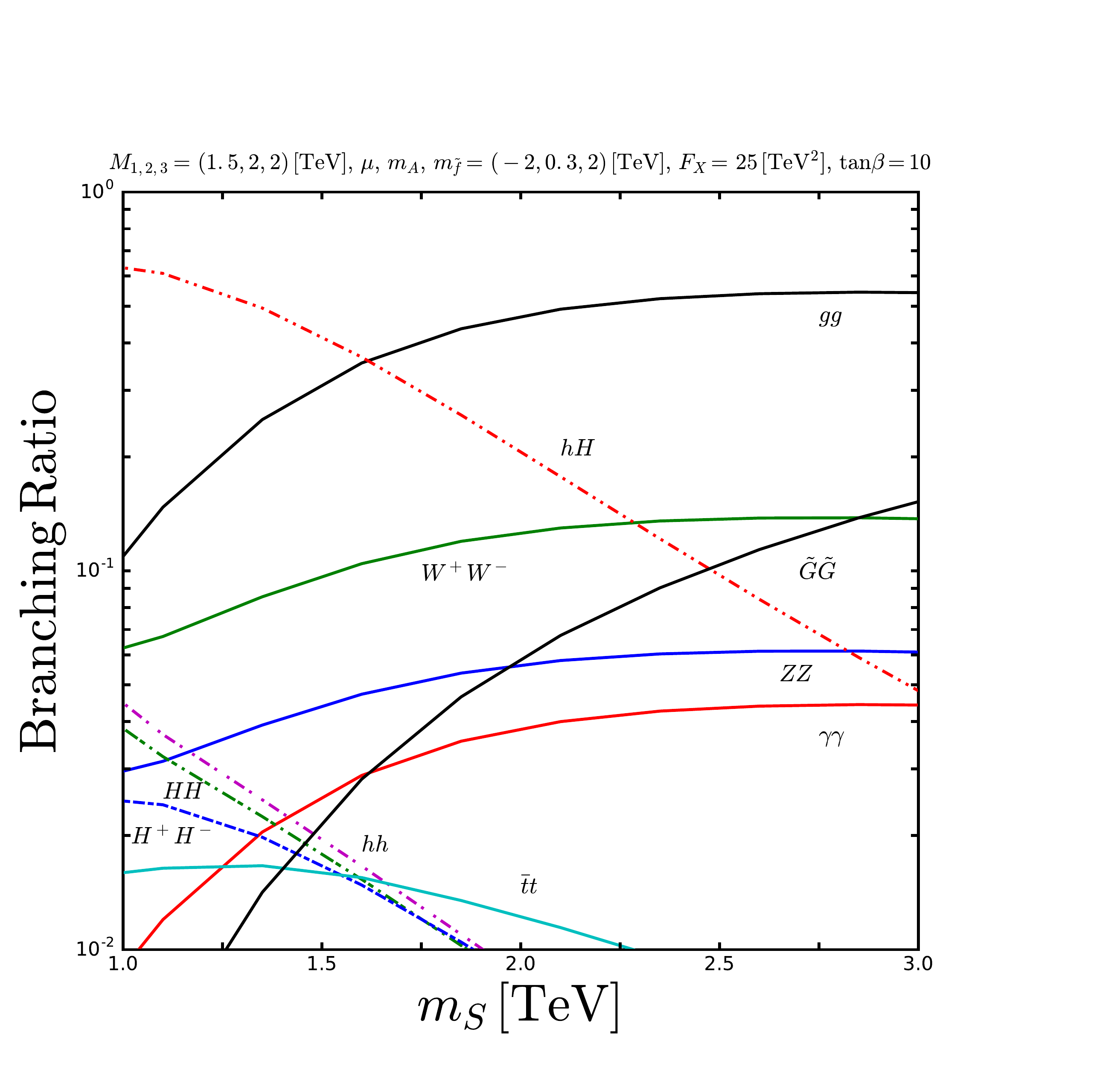}
  \includegraphics[width=7.5cm]{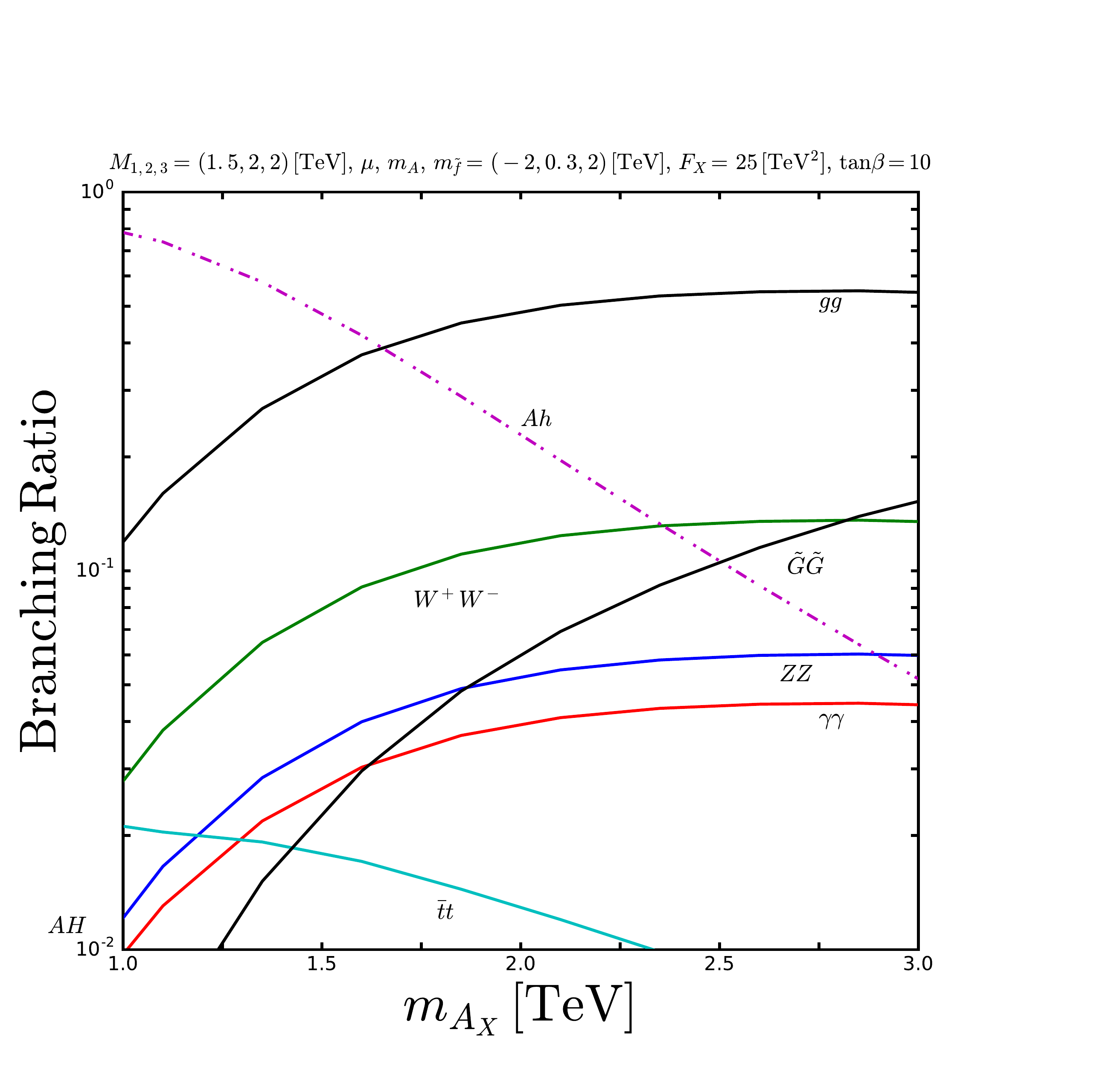}
 \end{center}
\caption{ Sample Point V: branching ratio of the sgoldstino (left panel) and pseudo-sgoldstino (right panel) at $\sqrt{F} = 5$ TeV, $(\mu, m_A, m_{\tilde{f}})=(-2,0.3,2)$ TeV and $(M_3,M_2,M_1) = (2, 2, 1.5)$ TeV with $\tan\beta=10$. 
}  
\label{fig:Bf_smallmA}
\end{figure}

%%%%%%%%%%%%%%%%%
Finally, we present results for sample point V in Fig.~\ref{fig:Bf_smallmA}.
The case of small $m_A$, where $m_A$ is $0.3$ TeV instead of $4$ TeV in sample point I.
Unlike the $hh$ decay mode of sgoldstino, the decay width of $hH$ ($hA$) is not $\tan\beta$ suppressed and depends on $\mu^2(m_A^2-2\mu^2)^2/(m_{\phi} F^2)$ as shown in Eq.~\eqref{eq:Width_hH_hA_app}. 
%Furthermore, it is not $v/\sqrt{F}$ suppressed. 
Thus, branching to $hH$ ($hA$) can be large if kinematically open. 
%
%===========================>
%===========================<

To summarize, the total decay width is not very large for the sample points considered here. If sgoldstino-Higgs mixing is not large, the total width can be extracted from each of the above figures using the approximate analytical expression for the width of $s \to gg$,

%--------------------------->
\begin{eqnarray}
\Gamma(s \to gg) 
&\approx& M_3^2 m_s^3/(4 \pi F^2)
\nonumber \\
&\sim& 0.5 {\rm GeV} \left( \frac{ m_s }{1 {\rm TeV}} \right)^3
                    \left( \frac{ M_3 }{2 {\rm TeV}} \right)^2
                    \left( \frac{ 5 {\rm TeV} }{ \sqrt{F} } \right)^4. 
  \label{eq:sgg_DecayWidth}
\end{eqnarray}
%---------------------------<
Thus, in the parameter space considered here, the total decay width is smaller than $100$ GeV and it can be measured as a narrow resonance at collider experiments.

%%%%%%%%%%%%%%%%%%%%%%%%%%%%%%%%%%%%%%%%%%%%%%%%%%%%%%%%%%%%%%%%%%%%%%%%%%%% 
%  Summary  
%%%%%%%%%%%%%%%%%%%%%%%%%%%%%%%%%%%%%%%%%%%%%%%%%%%%%%%%%%%%%%%%%%%%%%%%%%%% 
\section{Summary}

In supersymmetric extentions of SM, low-scale breaking of SUSY is phenomenologically valid. 
One of the features of low-scale SUSY breaking is that the hidden sector can be accessible in collider experiments as the couplings between SM and hidden sector are not suppressed by a high-scale mass parameter. Furthermore, there are additional contributions to quartic coupling of the lightest Higgs boson with which we can obtain Higgs mass of $125$ GeV at tree level \cite{Antoniadis:2010hs,Petersson:2011in,Dudas:2012fa,Petersson:2012nv}. 

We have investigated the collider phenomenology of sgoldstino which is the scalar component of the goldstino superfield. We have considered various possible branches of sgoldstino and pseudo-sgoldstino decay in this paper, including that of Higgs bosons, sparticles and particles final state. 

We have shown that sgoldstino decays to $s \to hh$ and longitudinal modes of $WW$ and $ZZ$ can be large if the $\mu$ parameter is large. If allowed kinematically, the branching to $s \to hH$ can be larger than $s \to hh$.
Finally, we have also discussed other possible collider phenomenology in the low-scale SUSY breaking scenario. In this scenario, the gravitino is very light as $m_{3/2} \sim 6 \times 10^{-3} {\rm eV} (\sqrt{F}/ (5 {\rm TeV}))^2$ and they can appear in the final state of SUSY particle production events at the LHC. Furthermore, the gravitino production may also be possible. For example, the gravitino-gluino production would provide large missing $E_T$ events at the LHC although the current constraint is not strong~\cite{Maltoni:2015twa}.

%%%%%%%%%%%%%%%%%%%%%%%%%%%%%%%%%%%%%%%%%%%%%%%%%%%%%%%%%%%%%%%%%%%%%%%%%%%%%%%%%
% Acknowledgments 
%%%%%%%%%%%%%%%%%%%%%%%%%%%%%%%%%%%%%%%%%%%%%%%%%%%%%%%%%%%%%%%%%%%%%%%%%%%%%%%%%
\section*{Acknowledgments}

This work is supported by the German Research Foundation through TRR33 "The Dark Universe" [MA, RG] and the Helmholtz Alliance for Astroparticle Physics [RG].

\appendix
%%%%%%%%%%%%%%%%%%%%%%%%%%%%%%%%%%%%%%%%%%%%%%%%%%%%%%%%%%%%%%%%%%%%%%%%%%%%%%%%%
% Appendix 
%%%%%%%%%%%%%%%%%%%%%%%%%%%%%%%%%%%%%%%%%%%%%%%%%%%%%%%%%%%%%%%%%%%%%%%%%%%%%%%%%
\section*{Appendix A: Lagrangian (before mixing)} \label{ap:A}

Here, we write the interaction terms relevant for production and decay of sgoldstino and pseudo-sgoldstino  at the LHC. We present the leading ($\mathcal{O}(1/F)$) contributions to sgoldstino production and decay.

\subsection*{Couplings with $gg$, $\gamma\gamma$ and $\gamma Z$}
Sgoldstino interactions with $gg$, $\gamma\gamma$ and $\gamma Z$ is given by 
%--------------------------->
\begin{eqnarray}
 \mathcal{L}_s &\supset& 
  \left(C_{sgg}\right) s_X G^{\mu \nu} G_{\mu \nu}
 +\left(C_{s\gamma\gamma}\right) s_X F^{\mu \nu} F_{\mu \nu}
 +\left(C_{s\gamma Z}\right) s_X F^{\mu \nu} Z_{\mu \nu}
\nonumber \\ &&
 +\left(C_{agg}\right) a_X G^{\mu \nu} \tilde{G}_{\mu \nu}
 +\left(C_{a\gamma\gamma}\right) a_X F^{\mu \nu} \tilde{F}_{\mu \nu}
 +\left(C_{a\gamma Z}\right) a_X F^{\mu \nu} \tilde{Z}_{\mu \nu},
  \label{eq:L_eff_1}
\end{eqnarray}
%---------------------------<
%--------------------------->
\begin{eqnarray}
C_{sgg} &=& - C_{agg} = - \frac{1}{2 \sqrt{2}} \frac{M_3}{F}, 
\nonumber \\ 
C_{s\gamma\gamma} &=& - C_{a\gamma\gamma} 
                  = - \frac{1}{2 \sqrt{2}} \frac{1}{F} \left( c_W^2 M_1 + s_W^2 M_2 \right), 
\nonumber \\ 
C_{s\gamma Z} &=& - C_{a\gamma Z}
              = - \frac{1}{\sqrt{2}} \frac{1}{F} s_W c_W \left( - M_1 +  M_2 \right), 
\nonumber
  \label{eq:L_eff_1_coeff}
\end{eqnarray}
%---------------------------<
where $\tilde{F}_{\mu \nu}$ is a dual field strength, $\tilde{F}_{\mu \nu} = (1/2) \epsilon_{\mu \nu \rho \sigma} \tilde{F}^{\rho \sigma}$.
We neglect MSSM $HGG$ term contribution in this paper, as these are suppressed by a loop factor. Since this is small, they are comparable to $sGG$ when the above couplings $ M_a/F \sim 10^{-5} $(GeV)$^{-1}$. 

\subsection*{Couplings with $WW$ and $ZZ$}
The sgoldstino interactions with $WW$ and $ZZ$ are written as 
%--------------------------->
\begin{eqnarray}
 \mathcal{L}_s &\supset& 
  \left(C_{sWW_T}\right) s_X W^{\mu \nu} W_{\mu \nu}
 +\left(C_{aWW_T}\right) a_X W^{\mu \nu} \tilde{W}_{\mu \nu}
\nonumber \\ &&
 +\left(C_{sZZ_T}\right) s_X Z^{\mu \nu} Z_{\mu \nu}
 +\left(C_{aZZ_T}\right) a_X Z^{\mu \nu} \tilde{Z}_{\mu \nu},
  \label{eq:L_eff_2}
\end{eqnarray}
%---------------------------<
%--------------------------->
\begin{eqnarray}
C_{sWW_T} &=& - C_{aWW_T} = - \frac{1}{\sqrt{2}} \frac{M_2}{F}, 
\nonumber \\ 
C_{sZZ_T} &=& - C_{aZZ_T} 
          = - \frac{1}{2 \sqrt{2}} \frac{1}{F} \left( s_W^2 M_1 + c_W^2 M_2 \right). 
\nonumber
  \label{eq:L_eff_2_coeff}
\end{eqnarray}
%---------------------------<
The interactions with longitudinal mode, e.g. $\left(C_{sWW_L}\right) m_W^2 s_X W^\mu W_\mu$, are $\mathcal{O}(1/F^2)$ terms. MSSM contributions which can affect the phenomenology of sgoldstino via mixing are 
%--------------------------->
\begin{eqnarray}
 \mathcal{L} &\supset& - 
  \left( \sin(\alpha-\beta) h -\cos(\alpha-\beta) H \right) 
   \left(  g_2 m_W W^{+\mu} W^-_\mu 
         + \frac{1}{2}\frac{g}{c_W} m_Z Z^\mu Z_\mu   \right) 
\\ &=& 
 C_{hWW_L} m_W^2 h W^{+\mu} W^-_\mu 
+C_{HWW_L} m_W^2 H W^{+\mu} W^-_\mu 
+C_{hZZ_L} m_Z^2 h Z^\mu Z_\mu 
+C_{HZZ_L} m_Z^2 H Z^\mu Z_\mu. 
\nonumber
  \label{eq:L_eff_2_MSSM}
\end{eqnarray}
%---------------------------<
%
%
%
%
%

\subsection*{Couplings with Higgs bosons}
The sgoldstino interactions with Higgs bosons are obtained as 
%--------------------------->
\begin{eqnarray}
 \mathcal{L}_s &\supset& 
  \left(C_{shh}\right) s_X h h
 +\left(C_{sHH}\right) s_X H H 
 +\left(C_{shH}\right) s_X h H 
 +\left(C_{sAA}\right) s_X A A
 +\left(C_{sH^+H^-}\right) s_X H^+ H^-
\nonumber \\ &&
 +\left(C_{ahA}\right) a_X h A
 +\left(C_{aHA}\right) a_X H A ,
  \label{eq:L_eff_3}
\end{eqnarray}
%---------------------------<
%--------------------------->
\begin{eqnarray}
C_{shh} &=&  \frac{1}{2 \sqrt{2}F} \left[ 
           \mu ((m_A^2 - 2 \mu^2) \sin2\alpha + m_A^2 \sin2\beta)
\right .\ \nonumber \\ && \qquad \qquad \qquad \left.\ 
         + m_Z^2 (s_W^2 M_1 + c_W^2 M_2) (1 - 2 \cos2\alpha \cos2\beta + \sin2\alpha \sin2\beta)  \right],
\nonumber \\ 
C_{sHH} &=& \frac{1}{2 \sqrt{2}F} \left[ 
           \mu (-(m_A^2 - 2 \mu^2) \sin2\alpha + m_A^2 \sin2\beta)
\right .\ \nonumber \\ && \qquad \qquad \qquad \left.\ 
         + m_Z^2 (s_W^2 M_1 + c_W^2 M_2) (1 + 2 \cos2\alpha \cos2\beta - \sin2\alpha \sin2\beta)  \right],
\nonumber \\
C_{shH} &=& \frac{1}{2 \sqrt{2}F} \left[ 
         -2\mu (m_A^2 - 2 \mu^2) \cos2\alpha 
         -m_Z^2 (s_W^2 M_1 + c_W^2 M_2) (3 \sin2(\alpha + \beta) + \sin2(\alpha - \beta))  \right],
\nonumber \\
C_{sAA} &=&  \frac{1}{2\sqrt{2}F} \left[ 
         2\mu (m_A^2 - \mu^2) \sin2\beta
         -m_Z^2 (s_W^2 M_1 + c_W^2 M_2) (\cos2\beta)^2  \right],
\nonumber \\ 
C_{sH^+H^-} &=&  \frac{1}{\sqrt{2}F} \left[ 
          2 \mu (m_A^2 - \mu^2) \sin2\beta 
         - m_Z^2 (s_W^2 M_1 (\cos2\beta)^2 - c_W^2 M_2 (1 + (\sin2\beta)^2))  \right],
\nonumber \\ 
C_{ahA} &=& \frac{1}{\sqrt{2} F} \left[ \mu (m_A^2 - 2 \mu^2) \sin(\alpha - \beta) \right],
\nonumber \\ 
C_{aHA} &=& \frac{1}{\sqrt{2} F} \left[ \mu (-m_A^2 + 2 \mu^2) \cos(\alpha - \beta) \right],
\nonumber 
  \label{eq:L_eff_3_coeff}
\end{eqnarray}
%---------------------------<
and the MSSM contributions which can affect sgoldstino phenomenology via mixing are
%--------------------------->
\begin{eqnarray}
 \mathcal{L} &\supset&
  -\frac{1}{4} \frac{g}{c_W} m_Z \cos2\alpha \sin(\alpha+\beta) h h h
%\nonumber \\ && 
  -\frac{1}{4} \frac{g}{c_W} m_Z \cos2\alpha \cos(\alpha+\beta) H H H
\nonumber \\ && 
  -\frac{1}{4} \frac{g}{c_W} m_Z \left( 2 \sin2\alpha \sin(\alpha+\beta) - \cos2\alpha \cos(\alpha+\beta) \right) h h H
\nonumber \\ && 
  +\frac{1}{4} \frac{g}{c_W} m_Z \left( 2 \sin2\alpha \cos(\alpha+\beta) + \cos2\alpha \sin(\alpha+\beta) \right) h H H
\nonumber \\ && 
  -\frac{1}{4} \frac{g}{c_W} m_Z \cos2\beta \sin(\alpha+\beta) h A A
%\nonumber \\ && 
  +\frac{1}{4} \frac{g}{c_W} m_Z \cos2\beta \cos(\alpha+\beta) H A A 
\nonumber \\ && 
  +\left( g m_W \sin(\alpha-\beta) - \frac{1}{2} \frac{g}{c_W} m_Z \cos2\beta \sin(\alpha+\beta)
   \right)  h H^+ H^- 
\nonumber \\ && 
  +\left(-g m_W \cos(\alpha-\beta) + \frac{1}{2} \frac{g}{c_W} m_Z \cos2\beta \cos(\alpha+\beta)
   \right)  H H^+ H^-. 
\nonumber \\ &=&
+C_{h h h} h h h
%\nonumber \\ && 
+C_{H H H} H H H
%\nonumber \\ &&  
+C_{h h H}  h h H
%\nonumber \\ &&  
+C_{h H H}  h H H
\nonumber \\ &&  
+C_{h A A}  h A A
%\nonumber \\ &&  
+C_{H A A}  H A A 
%\nonumber \\ &&  
+C_{h H^+ H^-}  h H^+ H^- 
%\nonumber \\ &&  
+C_{H H^+ H^-}  H H^+ H^-. 
  \label{eq:L_eff_3_MSSM}
\end{eqnarray}
%---------------------------<

\subsection*{Couplings with gauginos, Higgsinos and Goldstinos}
The sgoldstino (and neutral Higgs bosons) interactions with gauginos, Higgsinos and Goldstinos are 
%--------------------------->
\begin{eqnarray}
 \mathcal{L}_s &\supset& 
  \left(C_{s\psi_X\psi_X}\right) s_X \psi_X \psi_X 
 +\left(C_{a\psi_X\psi_X}\right)ia_X \psi_X \psi_X 
\nonumber \\ &&
 +\left(C_{s\tilde{V}\tilde{V}}^K \right) s_X \tilde{V} 
      \left( i \frac{\sigma^\mu}{2} \partial_\mu \right) \bar{\tilde{V}} 
 +\left(C_{a\tilde{V}\tilde{V}}^K \right) i a_X \tilde{V} 
      \left( i \frac{\sigma^\mu}{2} \partial_\mu \right) \bar{\tilde{V}} 
\nonumber \\ &&
 +\left(C_{h\psi_X\tilde{B}}\right) h \psi_X \tilde{B}
 +\left(C_{H\psi_X\tilde{B}}\right) H \psi_X \tilde{B}
%\nonumber \\ &&
 +\left(C_{h\psi_X\tilde{W}}\right) h \psi_X \tilde{W}^0
 +\left(C_{H\psi_X\tilde{W}}\right) H \psi_X \tilde{W}^0
\nonumber \\ &&
+\left(C_{h\Psi_X\tilde{H}_d^0}\right) h\Psi_X\tilde{H}_d^0
+\left(C_{h\Psi_X\tilde{H}_u^0}\right) h\Psi_X\tilde{H}_u^0 
+\left(C_{H\Psi_X\tilde{H}_d^0}\right) H\Psi_X\tilde{H}_d^0 
+\left(C_{H\Psi_X\tilde{H}_u^0}\right) H\Psi_X\tilde{H}_u^0
\nonumber \\ &&
+\left(C_{A\Psi_X\tilde{H}_d^0}\right)iA\Psi_X\tilde{H}_d^0
+\left(C_{A\Psi_X\tilde{H}_u^0}\right)iA\Psi_X\tilde{H}_u^0 
+\left(C_{s\tilde{H}_d^0\tilde{H}_u^0}\right) s_X \tilde{H}_d^0\tilde{H}_u^0 
+\left(C_{a\tilde{H}_d^0\tilde{H}_u^0}\right)ia_X \tilde{H}_d^0\tilde{H}_u^0 
\nonumber \\ &&
+\left(C_{s\tilde{H}_u^+\tilde{H}_d^-}\right) s_X \tilde{H}_u^+\tilde{H}_d^- 
+\left(C_{a\tilde{H}_u^+\tilde{H}_d^-}\right)ia_X \tilde{H}_u^+\tilde{H}_d^- + {\rm h.c.}, 
  \label{eq:L_eff_4}
\end{eqnarray}
%---------------------------<
%--------------------------->
\begin{eqnarray}
C_{s\psi_X\psi_X} &=& - \frac{1}{2 \sqrt{2}} \frac{m_X^2}{F} = - C_{a\psi_X\psi_X}, \qquad
C_{s\tilde{V}\tilde{V}}^K = \frac{\sqrt{2} M_a}{F}= C_{a\tilde{V}\tilde{V}}^K,  \qquad 
\nonumber \\ 
C_{h\psi_X\tilde{B}} &=& s_W m_Z M_1 \sin(\alpha+\beta) /(2 \sqrt{2} F), \qquad
%\nonumber \\ 
C_{H\psi_X\tilde{B}}  = -s_W m_Z M_1 \cos(\alpha+\beta) /(2 \sqrt{2} F),
\nonumber \\ 
C_{h\psi_X\tilde{W}} &=&-c_W m_Z M_2 \sin(\alpha+\beta) /(2 \sqrt{2} F), \qquad
%\nonumber \\ 
C_{H\psi_X\tilde{W}}  = c_W m_Z M_2 \cos(\alpha+\beta) /(2 \sqrt{2} F),
%\nonumber \\  
\nonumber \\ 
C_{h\Psi_X\tilde{H}_d^0} &=& 
  \left[ 2 m_A^2 \cos(\alpha - \beta) \sin\beta - (2 \mu^2 + m_Z^2 \cos2\beta) \sin\alpha  
  \right]/(2 \sqrt{2} F),
\nonumber \\ 
C_{h\Psi_X\tilde{H}_u^0} &=& 
  \left[ -2 m_A^2 \cos(\alpha - \beta) \cos\beta + (2 \mu^2 - m_Z^2 \cos2\beta) \cos\alpha 
  \right]/(2 \sqrt{2} F),
\nonumber \\ 
C_{H\Psi_X\tilde{H}_d^0} &=& 
  \left[ 2 m_A^2 \sin(\alpha - \beta) \sin\beta + (2 \mu^2 + m_Z^2 \cos2\beta) \cos\alpha 
  \right]/(2 \sqrt{2} F),
\nonumber \\ 
C_{H\Psi_X\tilde{H}_u^0} &=& 
  \left[ -2 m_A^2 \sin(\alpha - \beta) \cos\beta + (2 \mu^2 - m_Z^2 \cos2\beta) \sin\alpha 
  \right]/(2 \sqrt{2} F),
\nonumber \\ 
C_{A\Psi_X\tilde{H}_d^0} &=& 
  ( 2 m_A^2 - 2 \mu^2 - m_Z^2 \cos2\beta) \sin\beta /(2 \sqrt{2} F),
\nonumber \\ 
C_{A\Psi_X\tilde{H}_u^0} &=& 
  ( 2 m_A^2 - 2 \mu^2 + m_Z^2 \cos2\beta) \cos\beta /(2 \sqrt{2} F),
\nonumber \\ 
C_{s\tilde{H}_d^0\tilde{H}_u^0} &=& C_{a\tilde{H}_d^0\tilde{H}_u^0} 
                                 =  m_A^2 \sin2\beta / (2 \sqrt{2} F), 
\nonumber \\ 
C_{s\tilde{H}_u^+\tilde{H}_d^-} &=& C_{a\tilde{H}_u^+\tilde{H}_d^-} 
                                 = - m_A^2 \sin2\beta / (2 \sqrt{2} F), 
  \label{eq:L_eff_4_coeff}
\end{eqnarray}
%---------------------------<
%
where ($\tilde{V}\tilde{V}$) denotes 
($\tilde{B}\tilde{B}$),
($\tilde{W}^0\tilde{W}^0$),
($\tilde{W}^+\tilde{W}^-$) 
and 
($\lambda_{\tilde{g}}^a \lambda_{\tilde{g}}^a$), and $\lambda_{\tilde{g}}^a$ is the two-component gluino field.

Corresponding MSSM couplings are 
%--------------------------->
\begin{eqnarray}
 \mathcal{L} &\supset&
  +\frac{g s_W}{2 c_W} \left( -\sin\alpha h +\cos\alpha H -i\sin\beta A 
                      \right) \tilde{B} \tilde{H}_d^0
%\nonumber \\ && 
  -\frac{g s_W}{2 c_W} \left( \cos\alpha h +\sin\alpha H -i\cos\beta A 
                      \right) \tilde{B} \tilde{H}_u^0
\nonumber \\ && 
  -\frac{g}{2} \left( -\sin\alpha h +\cos\alpha H -i\sin\beta A 
                      \right) \tilde{W} \tilde{H}_d^0
%\nonumber \\ && 
  +\frac{g}{2} \left( \cos\alpha h +\sin\alpha H -i\cos\beta A 
                      \right) \tilde{W} \tilde{H}_u^0
\nonumber \\ && 
  -\frac{g}{\sqrt{2}} \left( -\sin\alpha h +\cos\alpha H -i\sin\beta A 
                      \right) \tilde{W}^+ \tilde{H}_d^-
\nonumber \\ && 
  -\frac{g}{\sqrt{2}} \left( \cos\alpha h +\sin\alpha H -i\cos\beta A
                      \right) \tilde{H}_u^+ \tilde{W}^-  + {\rm h. c. }
\nonumber \\ &=&
 \left( C_{h\tilde{B}\tilde{H}_d} h + C_{H\tilde{B}\tilde{H}_d} H + C_{A\tilde{B}\tilde{H}_d}iA 
 \right) \tilde{B} \tilde{H}_d^0
%\nonumber \\ && 
+\left( C_{h\tilde{B}\tilde{H}_u} h + C_{H\tilde{B}\tilde{H}_u} H + C_{A\tilde{B}\tilde{H}_u}iA 
 \right) \tilde{B} \tilde{H}_u^0 
\nonumber \\ && 
+\left( C_{h\tilde{W}\tilde{H}_d} h + C_{H\tilde{W}\tilde{H}_d} H + C_{A\tilde{W}\tilde{H}_d}iA 
 \right) \tilde{W} \tilde{H}_d^0
%\nonumber \\ && 
+\left( C_{h\tilde{W}\tilde{H}_u} h + C_{H\tilde{W}\tilde{H}_u} H + C_{A\tilde{W}\tilde{H}_u}iA 
 \right) \tilde{W} \tilde{H}_u^0 
\nonumber \\ && 
+\left( C_{h\tilde{W}^+\tilde{H}_d^-} h + C_{H\tilde{W}^+\tilde{H}_d^-} H 
       +C_{A\tilde{W}^+\tilde{H}_d^-}iA \right) \tilde{W}^+ \tilde{H}_d^-
\nonumber \\ && 
+\left( C_{h\tilde{H}_u^+\tilde{W}^-} h + C_{H\tilde{H}_u^+\tilde{W}^-} H 
       +C_{A\tilde{H}_u^+\tilde{W}^-}iA \right) \tilde{H}_u^+ \tilde{W}^- + {\rm h. c. }. 
  \label{eq:L_eff_4_MSSM}
\end{eqnarray}
%---------------------------<

\subsection*{Couplings with fermion and sfermions}

The mass matrices of sfermions are the same as in MSSM: 
%--------------------------->
\begin{eqnarray}
V &\supset& 
\left( \tilde{f}_L^* \tilde{f}_R^* \right)
\left(
\begin{array}{cc}
       m_{\tilde{f}_{LL}}^2 & m_{\tilde{f}_{LR}}^2 \\ 
       m_{\tilde{f}_{LR}}^{2*} & m_{\tilde{f}_{RR}}^2 \\ 
\end{array}
\right)
\left(
\begin{array}{c}
   \tilde{f}_L \\
   \tilde{f}_R \\ 
\end{array}
\right), 
\nonumber \\
m_{\tilde{f}_{LL}}^2 &=& m_{\tilde{f}_L}^2 + m_f^2 
                        +m_Z^2 \cos2\beta \left( T_{3L} - Q s_W^2\right) 
\nonumber \\
m_{\tilde{f}_{RR}}^2 &=& m_{\tilde{f}_R}^2 + m_f^2 
                        +m_Z^2 \cos2\beta \left( Q s_W^2 \right) 
\nonumber \\
m_{\tilde{f}_{LR}}^2 &=& m_u (A_u/y_u + \mu / \tan\beta) 
%\nonumber \\
 \quad {\rm (or)} \quad m_d (A_d/y_d + \mu \tan\beta), 
  \label{eq:sfermion_matrix}
\end{eqnarray}
%---------------------------<
where $y_u, y_d$ are defined in Eq.~\eqref{eq:Lagrangian0}. 
We define mass eigenbasis $\tilde{f}_i = (\tilde{f}_1,\tilde{f}_2)^T$ with $m_{\tilde{f}_1} < m_{\tilde{f}_2}$ by 
%--------------------------->
\begin{eqnarray}
\tilde{f}_i
&=& U^f \tilde{f}^\prime . 
  \label{eq:sfermion_rotation}
\end{eqnarray}
%---------------------------<
where $\tilde{f}^\prime = (\tilde{f}_L,\tilde{f}_R)$. 

Sgoldstino interactions with sfermions are given by, 
%--------------------------->
\begin{eqnarray}
 \mathcal{L}_s &\supset& 
 \left[ 
  \left(C_{sff}\right) s_X f_L f_R^c
 +\left(C_{aff}\right)ia_X f_L f_R^c
   +h.c. \right]
\nonumber \\ && 
 +\left(C_{s\tilde{t}_L\tilde{t}_L} \right) s_X \tilde{t}_L^* \tilde{t}_L 
 +\left(C_{s\tilde{t}_R\tilde{t}_R} \right) s_X \tilde{t}_R^* \tilde{t}_R 
\nonumber \\ && 
 + \left[ \left(C_{s\tilde{t}_L\tilde{t}_R} \right) s_X \tilde{t}_L^* \tilde{t}_R 
         +\left(C_{a\tilde{t}_L\tilde{t}_R} \right)ia_X \tilde{t}_L^* \tilde{t}_R 
   +h.c. \right]
  \label{eq:L_eff_5}
\end{eqnarray}
%---------------------------<
%--------------------------->
\begin{eqnarray}
C_{stt} &=& C_{att} = - \frac{1}{\sqrt{2} F} m_t \frac{A_t}{y_t}, 
\nonumber \\
C_{s\tilde{t}_L\tilde{t}_L} &=& 
 -\frac{1}{F} 
    \left[ -\sqrt{2} m_t^2 \frac{A_t}{y_t} + \sqrt{2} m_Z^2 (s_W^2 M_1 + c_W^2 M_2) 
                          \cos2\beta (T_3 - Q s_W^2) \right], 
\nonumber \\
C_{s\tilde{t}_R\tilde{t}_R} &=&
 -\frac{1}{F} 
    \left[ -\sqrt{2} m_t^2 \frac{A_t}{y_t} + \sqrt{2} m_Z^2 (s_W^2 M_1 + c_W^2 M_2) \cos2\beta (Q s_W^2) \right], 
\nonumber \\
C_{s\tilde{t}_L\tilde{t}_R} &=& - 
  \frac{1}{\sqrt{2} F} \frac{m_t}{\tan\beta}
           \left[ A_t \mu + \frac{1}{2} m_A^2 \sin2\beta \right] 
= - C_{a\tilde{t}_L\tilde{t}_R} , 
\nonumber 
  \label{eq:L_eff_5_coeff_t}
\end{eqnarray}
%---------------------------<
%--------------------------->
\begin{eqnarray}
C_{sbb} &=& C_{abb} = - \frac{1}{\sqrt{2} F} m_b \frac{A_b}{y_b} , 
\nonumber \\
C_{s\tilde{b}_L\tilde{b}_L} &=& 
 -\frac{1}{F} 
    \left[- \sqrt{2} m_b^2\frac{A_b}{y_b} + \sqrt{2} m_Z^2 (s_W^2 M_1 + c_W^2 M_2) 
                          \cos2\beta (T_3 - Q s_W^2) \right], 
\nonumber \\
C_{s\tilde{b}_R\tilde{b}_R} &=&
 -\frac{1}{F} 
    \left[- \sqrt{2} m_b^2\frac{A_b}{y_b} + \sqrt{2} m_Z^2 (s_W^2 M_1 + c_W^2 M_2) \cos2\beta (Q s_W^2) \right], 
\nonumber \\
C_{s\tilde{b}_L\tilde{b}_R} &=& -
  \frac{1}{\sqrt{2} F} m_b \tan\beta
           \left[ A_b \mu + \frac{1}{2} m_A^2 \sin2\beta \right]
= - C_{a\tilde{b}_L\tilde{b}_R} , 
\nonumber 
  \label{eq:L_eff_5_coeff_b}
\end{eqnarray}

The MSSM interactions are 
%--------------------------->
\begin{eqnarray}
 \mathcal{L}_s &\supset& 
  \left[
        \left( C_{hff} \right) h f_L f_R^c 
       +\left( C_{Hff} \right) H f_L f_R^c
       +\left( C_{Aff} \right)iA f_L f_R^c 
        + {\rm h.c.} \right]
\nonumber \\ && 
 +\left( C_{h\tilde{f}_L\tilde{f}_L} \right) h \tilde{f}_L^* \tilde{f}_L 
 +\left( C_{H\tilde{f}_L\tilde{f}_L} \right) H \tilde{f}_L^* \tilde{f}_L 
 +\left( C_{h\tilde{f}_R\tilde{f}_R} \right) h \tilde{f}_R^* \tilde{f}_R 
 +\left( C_{H\tilde{f}_R\tilde{f}_R} \right) H \tilde{f}_R^* \tilde{f}_R 
\nonumber \\ && 
 +\left[ 
        \left( C_{h\tilde{f}_L\tilde{f}_R} \right) h \tilde{f}_L^* \tilde{f}_R 
       +\left( C_{H\tilde{f}_L\tilde{f}_R} \right) H \tilde{f}_L^* \tilde{f}_R 
       +\left( C_{A\tilde{f}_L\tilde{f}_R} \right)iA \tilde{f}_L^* \tilde{f}_R  
   + h.c. \right],
  \label{eq:L_eff_5_MSSM}
\end{eqnarray}
%---------------------------<
%--------------------------->
\begin{eqnarray}
C_{htt} &=& - \frac{g m_t \cos\alpha}{2m_W \sin\beta}, \qquad
C_{Htt}  = - \frac{g m_t \sin\alpha}{2m_W \sin\beta}, \qquad 
C_{Att}  = - \frac{g m_t}{2m_W \tan\beta},
\nonumber \\ 
C_{h\tilde{t}_L\tilde{t}_L} &=& \frac{g m_t^2 \cos\alpha}{m_W \sin\beta} 
   - \frac{g}{c_W} m_Z \sin(\alpha+\beta) \left(T_3 - Q s_W^2 \right), 
\nonumber \\
C_{h\tilde{t}_R\tilde{t}_R} &=& \frac{g m_t^2 \cos\alpha}{m_W \sin\beta} 
   - \frac{g}{c_W} m_Z \sin(\alpha+\beta) \left(Q s_W^2 \right), 
\nonumber \\
C_{h\tilde{t}_L\tilde{t}_R} &=& \frac{g m_t}{2 m_W \sin\beta} 
     \left( A_t/y_t \cos\alpha - \mu \sin\alpha  \right), 
\nonumber \\ 
C_{H\tilde{t}_L\tilde{t}_L} &=& \frac{g m_t^2 \sin\alpha}{m_W \sin\beta} 
   + \frac{g}{c_W} m_Z \cos(\alpha+\beta) \left(T_3 - Q s_W^2 \right), 
\nonumber \\ 
C_{H\tilde{t}_R\tilde{t}_R} &=& \frac{g m_t^2 \sin\alpha}{m_W \sin\beta} 
   + \frac{g}{c_W} m_Z \cos(\alpha+\beta) \left( Q s_W^2 \right), 
\nonumber \\ 
C_{H\tilde{t}_L\tilde{t}_R} &=& \frac{g m_t}{2 m_W \sin\beta} 
     \left( A_t/y_t \sin\alpha + \mu \cos\alpha  \right), 
\nonumber \\ 
C_{A\tilde{t}_L\tilde{t}_R} &=& -\frac{g m_t}{2 m_W} 
     \left( -(A_t/y_t)/\tan\beta + \mu \right),
\nonumber
  \label{eq:L_eff_5_MSSM_coeff_t}
\end{eqnarray}
%---------------------------<
%--------------------------->
\begin{eqnarray}
C_{hbb} &=& \frac{g m_b \sin\alpha}{2m_W \cos\beta}, \qquad
C_{Hbb}  = -\frac{g m_b \cos\alpha}{2m_W \cos\beta}, \qquad 
C_{Abb}  = -\frac{g m_b}{2m_W}\tan\beta,
\nonumber \\ 
C_{h\tilde{b}_L\tilde{b}_L} &=& -\frac{g m_b^2 \sin\alpha}{m_W \cos\beta} 
   - \frac{g}{c_W} m_Z \sin(\alpha+\beta) \left(T_3 - Q s_W^2 \right), 
\nonumber \\
C_{h\tilde{b}_R\tilde{b}_R} &=& -\frac{g m_b^2 \sin\alpha}{m_W \cos\beta} 
   - \frac{g}{c_W} m_Z \sin(\alpha+\beta) \left(Q s_W^2 \right), 
\nonumber \\
C_{h\tilde{b}_L\tilde{b}_R} &=& \frac{g m_b}{2 m_W \cos\beta} 
     \left( -A_b/y_b \sin\alpha + \mu \cos\alpha  \right), 
\nonumber \\ 
C_{H\tilde{b}_L\tilde{b}_L} &=& \frac{g m_b^2 \cos\alpha}{m_W \cos\beta} 
   + \frac{g}{c_W} m_Z \cos(\alpha+\beta) \left(T_3 - Q s_W^2 \right), 
\nonumber \\ 
C_{H\tilde{b}_R\tilde{b}_R} &=& \frac{g m_b^2 \cos\alpha}{m_W \cos\beta} 
   + \frac{g}{c_W} m_Z \cos(\alpha+\beta) \left( Q s_W^2 \right), 
\nonumber \\ 
C_{H\tilde{b}_L\tilde{b}_R} &=& \frac{g m_b}{2 m_W \cos\beta} 
     \left( A_b/y_b \cos\alpha + \mu \sin\alpha  \right), 
\nonumber \\ 
C_{A\tilde{b}_L\tilde{b}_R} &=& -\frac{g m_b}{2 m_W} 
     \left( -(A_b/y_b) \tan\beta + \mu \right).
\nonumber
  \label{eq:L_eff_5_MSSM_coeff_b}
\end{eqnarray}
%---------------------------<

%%%%%%%%%%%%%%%%%%%%%%%%%%%%%%%%%%%%%%%%%%%%%%%%%%%%%%%%%%%%%%%%%%%%%%%%%%%%%%%%%
% Appendix 
%%%%%%%%%%%%%%%%%%%%%%%%%%%%%%%%%%%%%%%%%%%%%%%%%%%%%%%%%%%%%%%%%%%%%%%%%%%%%%%%%
\section*{Appendix B: Lagrangian} \label{ap:B}

We now show the interaction terms written in the  mass basis. 
\subsection*{Couplings to $gg$, $\gamma \gamma$ and $\gamma Z$}
%--------------------------->
\begin{eqnarray}
 \mathcal{L}_s^{\rm gauge 1} &=& 
  \left(C_{\phi_i gg}\right) \phi_i  G^{\mu \nu} G_{\mu \nu}
 +\left(C_{\phi_i \gamma\gamma}\right) \phi_i  F^{\mu \nu} F_{\mu \nu}
 +\left(C_{\phi_i \gamma Z}\right) \phi_i  F^{\mu \nu} Z_{\mu \nu}
\nonumber \\ && 
 +\left(C_{\phi_{ai} gg}\right) a_i  G^{\mu \nu} \tilde{G}_{\mu \nu}
 +\left(C_{\phi_{ai} \gamma\gamma}\right) a_i  F^{\mu \nu} \tilde{F}_{\mu \nu}
 +\left(C_{\phi_{ai} \gamma Z}\right) a_i  F^{\mu \nu} \tilde{Z}_{\mu \nu},
  \label{eq:L_eff_1}
\end{eqnarray}
%---------------------------<
where
%--------------------------->
\begin{eqnarray}
C_{\phi_i gg} &=& \sum_j S_{ij} C_{h_j gg}, \qquad \quad
C_{\phi_{ai} gg} = \sum_j A_{ij} C_{A_j gg},
\nonumber \\ 
C_{\phi_i \gamma\gamma} &=& \sum_j S_{ij} C_{h_j \gamma\gamma}, \qquad \quad 
C_{\phi_{ai} \gamma\gamma} = \sum_j A_{ij} C_{A_j \gamma\gamma},
\nonumber \\ 
C_{\phi_i \gamma Z} &=& \sum_j S_{ij} C_{h_j \gamma Z}, \qquad \quad 
C_{\phi_{ai} \gamma Z} = \sum_j A_{ij} C_{A_j \gamma Z}. 
\nonumber
  \label{eq:L_eff_1_c}
\end{eqnarray}
%---------------------------<

\subsection*{Couplings to $WW$ and $ZZ$}
%--------------------------->
\begin{eqnarray}
 \mathcal{L}_s^{\rm gauge 2} &=& 
  \left(C_{\phi_i WW_T}\right) \phi_i  W^{\mu \nu} W_{\mu \nu}
 +\left(C_{\phi_i WW_L}\right) m_W^2 \phi_i  W^\mu W_\mu
\nonumber \\ &&
 +\left(C_{\phi_i ZZ_T}\right) \phi_i  Z^{\mu \nu} Z_{\mu \nu}
 +\left(C_{\phi_i ZZ_L}\right) m_Z^2 \phi_i  Z^\mu Z_\mu
\nonumber \\ &&
 +\left(C_{\phi_{ai} WW_T}\right) \phi_{ai}  W^{\mu \nu} \tilde{W}_{\mu \nu}
 +\left(C_{\phi_{ai} ZZ_T}\right) \phi_{ai}  Z^{\mu \nu} \tilde{Z}_{\mu \nu},
  \label{eq:L_eff_2}
\end{eqnarray}
%---------------------------<
where
%--------------------------->
\begin{eqnarray}
C_{\phi_i WW_T} &=& \sum_j S_{ij} C_{h_j WW_T}, \qquad \quad 
C_{\phi_{ai} WW_T} = \sum_j A_{ij} C_{A_j WW_T},
\nonumber \\
C_{\phi_i ZZ_T} &=& \sum_j S_{ij} C_{h_j ZZ_T}, \qquad \quad 
C_{\phi_{ai} ZZ_T} = \sum_j A_{ij} C_{A_j ZZ_T},
\nonumber \\
C_{\phi_i WW_L} &=& \sum_j S_{ij} C_{h_j WW_L}, 
\nonumber \\
C_{\phi_i ZZ_L} &=& \sum_j S_{ij} C_{h_j ZZ_L}. 
\nonumber
  \label{eq:L_eff_2_c}
\end{eqnarray}
%---------------------------<

\subsection*{Couplings to Higgs bosons}
%--------------------------->
\begin{eqnarray}
 \mathcal{L}_s^{\rm scalar} &=& 
  \left(C_{\phi_i\phi_j\phi_k}\right) \phi_i\phi_j\phi_k 
 +\left(C_{\phi_i \phi_{aj} \phi_{ak}} \right) \phi_i a_j a_k
 +\left(C_{\phi_i H^+H^-}\right) \phi_i H^+ H^-,
  \label{eq:L_eff_3}
\end{eqnarray}
%---------------------------<
where
%--------------------------->
\begin{eqnarray}
C_{\phi_i\phi_j\phi_k} &=& \sum_{i^\prime j^\prime k^\prime} 
                                S_{ii^\prime} S_{jj^\prime} S_{kk^\prime} 
                                C_{h_{i^\prime} h_{j^\prime} h_{k^\prime}}, 
\nonumber \\ 
C_{\phi_i\phi_{aj}\phi_{ak}} &=& \sum_{i^\prime j^\prime k^\prime} 
                                S_{ii^\prime} A_{jj^\prime} A_{kk^\prime} 
                                C_{h_{i^\prime} A_{j^\prime} A_{k^\prime}}, 
\nonumber \\ 
C_{\phi_i H^+H^-} &=& \sum_j S_{ij} C_{h_j H^+H^-}.
\nonumber
  \label{eq:L_eff_3_c}
\end{eqnarray}
%---------------------------<

\subsection*{Couplings to Neutralinos and Goldstinos}
%--------------------------->
\begin{eqnarray}
 \mathcal{L}_s^{\rm neutralino} &=& 
  \phi_i  \bar{\psi}_{\tilde{\chi}_j} 
  \left[\left(C_{\phi_i \psi_{\tilde{\chi}_j} \psi_{\tilde{\chi}_k}}^S\right) 
       +\left(C_{\phi_i \psi_{\tilde{\chi}_j} \psi_{\tilde{\chi}_k}}^P\right) i \gamma_5 
       +\left(C_{\phi_i \psi_{\tilde{\chi}_j} \psi_{\tilde{\chi}_k}}^{KS}\right) 
        i \frac{\Slash{\partial}}{2} 
       +\left(C_{\phi_i \psi_{\tilde{\chi}_j} \psi_{\tilde{\chi}_k}}^{KP}\right) 
          \frac{\Slash{\partial}}{2} \gamma_5 
\right] 
               \psi_{\tilde{\chi}_k} 
\nonumber \\ && 
 +a_i  \bar{\psi}_{\tilde{\chi}_j} 
  \left[\left(C_{\phi_{ai} \psi_{\tilde{\chi}_j} \psi_{\tilde{\chi}_k}}^S\right) 
       +\left(C_{\phi_{ai} \psi_{\tilde{\chi}_j} \psi_{\tilde{\chi}_k}}^P\right) i \gamma_5           \frac{1}{2} \overset{\leftrightarrow}{\Slash{\partial}} \gamma_5
       +\left(C_{\phi_{ai} \psi_{\tilde{\chi}_j} \psi_{\tilde{\chi}_k}}^{KS}\right) 
        i \frac{\Slash{\partial}}{2} 
       +\left(C_{\phi_{ai} \psi_{\tilde{\chi}_j} \psi_{\tilde{\chi}_k}}^{KP}\right) 
          \frac{\Slash{\partial}}{2} \gamma_5
\right] 
               \psi_{\tilde{\chi}_k}, 
\nonumber \\
&&
  \label{eq:L_eff_4}
\end{eqnarray}

%---------------------------<
where
%--------------------------->
\begin{eqnarray}
% phi S
C_{\phi_i \psi_{\tilde{\chi}_j} \psi_{\tilde{\chi}_k}}^S 
&=& \sum_{i^\prime j^\prime k^\prime} S_{ii^\prime} N^\prime_{jj^\prime} N^\prime_{kk^\prime} 
C_{h_{i^\prime} \tilde{N}_{j^\prime}^0 \tilde{N}_{k^\prime}^0} 
{\rm Re}(\xi_j \xi_k),
\nonumber \\ 
% phi P
C_{\phi_i \psi_{\tilde{\chi}_j} \psi_{\tilde{\chi}_k}}^P 
&=& \sum_{i^\prime j^\prime k^\prime} S_{ii^\prime} N^\prime_{jj^\prime} N^\prime_{kk^\prime} 
C_{h_{i^\prime} \tilde{N}_{j^\prime}^0 \tilde{N}_{k^\prime}^0} 
\{ - {\rm Im}(\xi_j \xi_k) \},
\nonumber \\ 
% phi KS
C_{\phi_i \psi_{\tilde{\chi}_j} \psi_{\tilde{\chi}_k}}^{KS} 
&=& \sum_{i^\prime j^\prime k^\prime} S_{ii^\prime} N^\prime_{jj^\prime} N^\prime_{kk^\prime} 
C_{h_{i^\prime} \tilde{N}_{j^\prime}^0 \tilde{N}_{k^\prime}^0}^K 
{\rm Re}(\xi_j^* \xi_k),
\nonumber \\ 
% phi KP
C_{\phi_i \psi_{\tilde{\chi}_j} \psi_{\tilde{\chi}_k}}^{KP} 
&=& \sum_{i^\prime j^\prime k^\prime} S_{ii^\prime} N^\prime_{jj^\prime} N^\prime_{kk^\prime} 
C_{h_{i^\prime} \tilde{N}_{j^\prime}^0 \tilde{N}_{k^\prime}^0}^K 
{\rm Im}(\xi_j^* \xi_k),
\nonumber \\ 
% a S
C_{\phi_{ai} \psi_{\tilde{\chi}_j} \psi_{\tilde{\chi}_k}}^S 
&=& \sum_{i^\prime j^\prime k^\prime} A_{ii^\prime} N^\prime_{jj^\prime} N^\prime_{kk^\prime} 
C_{A_{i^\prime} \tilde{N}_{j^\prime}^0 \tilde{N}_{k^\prime}^0} 
\{ - {\rm Im}(\xi_j \xi_k) \},
\nonumber \\ 
% a P
C_{\phi_{ai} \psi_{\tilde{\chi}_j} \psi_{\tilde{\chi}_k}}^P 
&=& \sum_{i^\prime j^\prime k^\prime} A_{ii^\prime} N^\prime_{jj^\prime} N^\prime_{kk^\prime} 
C_{A_{i^\prime} \tilde{N}_{j^\prime}^0 \tilde{N}_{k^\prime}^0} 
\{ - {\rm Re}(\xi_j \xi_k) \},
\nonumber \\ 
% a KS 
C_{\phi_{ai} \psi_{\tilde{\chi}_j} \psi_{\tilde{\chi}_k}}^{KS} 
&=& \sum_{i^\prime j^\prime k^\prime} A_{ii^\prime} N^\prime_{jj^\prime} N^\prime_{kk^\prime} 
C_{A_{i^\prime} \tilde{N}_{j^\prime}^0 \tilde{N}_{k^\prime}^0}^K 
{\rm Im}(\xi_j^* \xi_k), 
%C_{h_{i^\prime} \psi_{j^\prime} \psi_{k^\prime}} \{ - {\rm Re}(\xi_j \xi_k) \},
\nonumber \\ 
% a KP 
C_{\phi_{ai} \psi_{\tilde{\chi}_j} \psi_{\tilde{\chi}_k}}^{KP} 
&=& \sum_{i^\prime j^\prime k^\prime} A_{ii^\prime} N^\prime_{jj^\prime} N^\prime_{kk^\prime} 
C_{A_{i^\prime} \tilde{N}_{j^\prime}^0 \tilde{N}_{k^\prime}^0}^K 
\{ -{\rm Re}(\xi_j^* \xi_k) \}. 
%C_{h_{i^\prime} \psi_{j^\prime} \psi_{k^\prime}} \{ - {\rm Re}(\xi_j \xi_k) \},
%% a K
%C_{\phi_{ai} \psi_{\tilde{\chi}_j}^0 \psi_{\tilde{\chi}_k}^0}^K 
%&=& \sum_{i^\prime j^\prime k^\prime} A_{ii^\prime} N_{jj^\prime} N_{kk^\prime} 
%C_{A_{i^\prime} \tilde{N}_{j^\prime} \tilde{N}_{k^\prime}}^K \{ - 1 \}. 
%%C_{h_{i^\prime} \psi_{j^\prime} \psi_{k^\prime}} \{ - {\rm Re}(\xi_j \xi_k) \},
%%\nonumber \\ 
\nonumber
  \label{eq:L_eff_4_c}
\end{eqnarray}
%---------------------------<

\subsection*{Couplings to Charginos}

\begin{eqnarray}
\mathcal{L}_s^{\rm chargino}&=& 
\phi_i\bar{\psi}_{\tilde{\chi}_j^+}\left[
 \left(C^S_{\phi_i \psi_{\tilde{\chi}_j^+} \psi_{\tilde{\chi}_k^-} } \right) 
+\left(C^P_{\phi_i \psi_{\tilde{\chi}_j^+} \psi_{\tilde{\chi}_k^-} } \right)\gamma_5 
\right . \nonumber \\ 
&& \left . \qquad \qquad 
+\left(C^{KS}_{\phi_i \psi_{\tilde{\chi}_j^+} \psi_{\tilde{\chi}_k^-} } \right) 
   i \frac{\overrightarrow{\Slash{\partial}} - \overleftarrow{\Slash{\partial}} }{2} 
+\left(C^{KP}_{\phi_i \psi_{\tilde{\chi}_j^+} \psi_{\tilde{\chi}_k^-} } \right) 
   i \frac{\overrightarrow{\Slash{\partial}} - \overleftarrow{\Slash{\partial}} }{2} \gamma_5 
                                  \right] \psi_{\tilde{\chi}_k^-} 
\nonumber \\ &&
+a_i\bar{\psi}_{\tilde{\chi}_j^+}\left[
 \left(C^S_{a_i \psi_{\tilde{\chi}_j^+} \psi_{\tilde{\chi}_k^-} } \right) i 
+\left(C^P_{a_i \psi_{\tilde{\chi}_j^+} \psi_{\tilde{\chi}_k^-} } \right) i \gamma_5 
\right . \nonumber \\ 
&& \left . \qquad \qquad 
+\left(C^{KS}_{a_i \psi_{\tilde{\chi}_j^+} \psi_{\tilde{\chi}_k^-} } \right) 
    \frac{\overrightarrow{\Slash{\partial}} + \overleftarrow{\Slash{\partial}} }{2} 
+\left(C^{KP}_{a_i \psi_{\tilde{\chi}_j^+} \psi_{\tilde{\chi}_k^-} } \right) 
    \frac{\overrightarrow{\Slash{\partial}} + \overleftarrow{\Slash{\partial}} }{2} \gamma_5
                                \right] \psi_{\tilde{\chi}_k^-}, \nonumber \\
&&
 \label{eq:L_eff_4chargino}
\end{eqnarray}
%---------------------------<
where
%--------------------------->
\begin{eqnarray}
C^S_{\phi_i \psi_{\tilde{\chi}_j}^+ \psi_{\tilde{\chi}_k}^-} 
&=& \sum_{i^\prime j^\prime k^\prime} \frac{1}{2} S_{ii^\prime} 
\left(C^R_{j j^\prime} C^L_{kk^\prime} + C^{R}_{kj^\prime} C^L_{jk^\prime} \right)
 C_{h_{i^\prime} \tilde{C}_{j^\prime}^+ \tilde{C}_{k^\prime}^-},
\nonumber \\ 
C^P_{\phi_i \psi_{\tilde{\chi}_j}^+ \psi_{\tilde{\chi}_k}^-} 
&=&\sum_{i^\prime j^\prime k^\prime} \frac{1}{2}S_{ii^\prime} 
\left(C^R_{j j^\prime} C^L_{kk^\prime} - C^{R}_{kj^\prime} C^L_{jk^\prime} \right)
 C_{h_{i^\prime} \tilde{C}_{j^\prime}^+ \tilde{C}_{k^\prime}^-},
\nonumber \\ 
C^{KS}_{\phi_i \psi_{\tilde{\chi}_j}^+ \psi_{\tilde{\chi}_k}^-} 
&=& \sum_{i^\prime j^\prime k^\prime} \frac{1}{2} S_{ii^\prime} 
\left(C^R_{j j^\prime} C^R_{kk^\prime} + C^L_{j j^\prime} C^L_{kk^\prime} \right)
 C_{h_{i^\prime} \tilde{C}_{j^\prime}^+ \tilde{C}_{k^\prime}^-}^K,
\nonumber \\ 
C^{KP}_{\phi_i \psi_{\tilde{\chi}_j}^+ \psi_{\tilde{\chi}_k}^-} 
&=& \sum_{i^\prime j^\prime k^\prime} \frac{1}{2} S_{ii^\prime} 
\left(C^R_{j j^\prime} C^R_{kk^\prime} - C^L_{j j^\prime} C^L_{kk^\prime} \right)
 C_{h_{i^\prime} \tilde{C}_{j^\prime}^+ \tilde{C}_{k^\prime}^-}^K,
\nonumber \\ 
C^S_{a_i \psi_{\tilde{\chi}_j}^+ \psi_{\tilde{\chi}_k}^-} 
&=& \sum_{i^\prime j^\prime k^\prime} \frac{1}{2} S_{ii^\prime} 
\left(- C^R_{jj^\prime} C^L_{kk^\prime} + C^{R}_{kj^\prime} C^L_{jk^\prime} \right)
 C_{A_{i^\prime} \tilde{C}_{j^\prime}^+ \tilde{C}_{k^\prime}^-},
\nonumber \\ 
C^P_{a_i \psi_{\tilde{\chi}_j}^+ \psi_{\tilde{\chi}_k}^-} 
&=&\sum_{i^\prime j^\prime k^\prime} \frac{1}{2} S_{ii^\prime} 
\left(- C^R_{jj^\prime} C^L_{kk^\prime} - C^{R}_{kj^\prime} C^L_{jk^\prime} \right)
 C_{A_{i^\prime} \tilde{C}_{j^\prime}^+ \tilde{C}_{k^\prime}^-},
\nonumber \\ 
C^{KS}_{a_i \psi_{\tilde{\chi}_j}^+ \psi_{\tilde{\chi}_k}^-} 
&=&\sum_{i^\prime j^\prime k^\prime} \frac{1}{2} S_{ii^\prime} 
\left(- C^R_{j j^\prime} C^R_{kk^\prime} + C^L_{j j^\prime} C^L_{kk^\prime} \right)
 C_{A_{i^\prime} \tilde{C}_{j^\prime}^+ \tilde{C}_{k^\prime}^-}^K, 
\nonumber \\ 
C^{KP}_{a_i \psi_{\tilde{\chi}_j}^+ \psi_{\tilde{\chi}_k}^-} 
&=&\sum_{i^\prime j^\prime k^\prime} \frac{1}{2} S_{ii^\prime} 
\left(- C^R_{j j^\prime} C^R_{kk^\prime} - C^L_{j j^\prime} C^L_{kk^\prime} \right)
 C_{A_{i^\prime} \tilde{C}_{j^\prime}^+ \tilde{C}_{k^\prime}^-}^K. 
\nonumber
  \label{eq:L_eff_4chargino_c}
\end{eqnarray}

\subsection*{Couplings to Gluino}
%--------------------------->
\begin{eqnarray}
 \mathcal{L}_s^{\rm ino} &=& 
\left(C_{\phi_i \tilde{g} \tilde{g}}^K \right) 
  \phi_i \bar{\psi}_{\tilde{g}}^a 
\left( i \frac{\Slash{\partial}}{2} \right)
               \psi_{\tilde{g}}^a 
+\left(C_{\phi_{ai} \tilde{g} \tilde{g}}^K \right) 
     a_i \bar{\psi}_{\tilde{g}}^a
\left( \frac{\Slash{\partial}}{2} \gamma_5 \right) 
               \psi_{\tilde{g}}^a, 
  \label{eq:L_eff_4p}
\end{eqnarray}
%---------------------------<
where
%--------------------------->
\begin{eqnarray}
% phi K
C_{\phi_i \tilde{g} \tilde{g}}^K
&=& \sum_{j} S_{ij} C_{h_j \tilde{g} \tilde{g}}^K, \qquad 
% a K
C_{\phi_{ai} \tilde{g} \tilde{g}}^K 
= \sum_{j} A_{ij} C_{A_j \tilde{g} \tilde{g}}^K \{ - 1 \}. 
\nonumber
  \label{eq:L_eff_4p_c}
\end{eqnarray}
%---------------------------<

\subsection*{Couplings to fermion and sfermions}
%--------------------------->
\begin{eqnarray}
 \mathcal{L}_s &\supset& 
 +\left(C_{\phi_i \tilde{f}_j \tilde{f}_k}\right) \phi_i \tilde{f}_j^* \tilde{f}_k
 +\left(C_{\phi_{ai} \tilde{f}_j \tilde{f}_k}\right) i a_i \tilde{f}_j^* \tilde{f}_k
 +\left(C_{\phi_i ff}\right) \phi_i \bar{\psi}_f \psi_f 
 +\left(C_{\phi_{ai} ff}\right) i a_i \bar{\psi}_f \gamma_5 \psi_f,
\nonumber \\
&&
  \label{eq:L_eff_5}
\end{eqnarray}
%---------------------------<
where
%--------------------------->
\begin{eqnarray}
C_{\phi_i \tilde{f}\tilde{f}} &=& 
\sum_{i^\prime j^\prime k^\prime} S_{ii^\prime} U_{jj^\prime}^{f*} U_{kk^\prime}^f C_{h_{i^\prime} \tilde{f}_{j^\prime} \tilde{f}_{k^\prime} }, \qquad
C_{\phi_{ai} \tilde{f}\tilde{f}} =
\sum_{i^\prime j^\prime k^\prime} A_{ii^\prime} U_{jj^\prime}^{f*} U_{kk^\prime}^f C_{A_{i^\prime} \tilde{f}_{j^\prime} \tilde{f}_{k^\prime} },
\nonumber \\ 
C_{\phi_i ff} &=& \sum_j S_{ij} C_{h_j ff},  \qquad  \qquad \qquad ~\quad
C_{\phi_{ai} ff} = - \sum_j A_{ij} C_{A_j ff}. 
\nonumber
  \label{eq:L_eff_5_c}
\end{eqnarray}
%---------------------------<

\section*{Appendix C: Decay width} \label{ap:C}

From the effective Lagrangian presented in Appendix B, the decay widths of $\phi_i$, which includes the sgoldstino, into SM gauge bosons and gravitino $\tilde{G}$ are obtained as
%--------------------------->
\begin{eqnarray}
 \Gamma(\phi_i \to gg) &=& \frac{2}{\pi} C_{\phi_i gg}^2 m_{\phi_i}^3,
\nonumber \\
 \Gamma(\phi_i \to \gamma \gamma) &=& \frac{1}{4\pi} C_{\phi_i \gamma\gamma}^2 m_{\phi_i}^3,
\nonumber \\
 \Gamma(\phi_i \to \gamma Z) &=& \frac{1}{8\pi} C_{\phi_i \gamma Z}^2 m_{\phi_i}^3 \left( 1-\frac{m_Z^2}{m_{\phi_i}^2} \right)^3,
\nonumber \\
 \Gamma(\phi_i \to WW) &=& \frac{1}{16\pi} \frac{m_W^4}{m_{\phi_i}} 
                       \left[2C_{\phi_i WW_T}^2 \left( 6 - 4 \frac{m_{\phi_i}^2}{m_W^2} + \frac{m_{\phi_i}^4}{m_W^4} \right) 
                             - 12 C_{\phi_i WW_T} C_{\phi_i WW_L} \left( 1 - \frac{m_{\phi_i}^2}{2 m_W^2} \right)
\right . \nonumber \\ 
&& \left . \qquad \qquad \qquad 
                             + C_{\phi_i WW_L}^2 \left( 3 - \frac{m_{\phi_i}^2}{m_W^2} + \frac{1}{4}\frac{m_{\phi_i}^4}{m_W^4} \right)
                      \right] \sqrt{1 - \frac{4 m_W^2}{m_{\phi_i}^2}}, 
\nonumber \\
 \Gamma(\phi_i \to ZZ) &=& \frac{1}{8\pi} \frac{m_Z^4}{m_{\phi_i}} 
                       \left[2 C_{\phi_i ZZ_T}^2 \left( 6 - 4 \frac{m_{\phi_i}^2}{m_Z^2} + \frac{m_{\phi_i}^4}{m_Z^4} \right) 
                             - 12 C_{\phi_i ZZ_T} C_{\phi_i ZZ_L} \left( 1 - \frac{m_{\phi_i}^2}{2 m_Z^2} \right)
\right . \nonumber \\ 
&& \left . \qquad \qquad \qquad 
                             + C_{\phi_i ZZ_L}^2 \left( 3 - \frac{m_{\phi_i}^2}{m_Z^2} + \frac{1}{4}\frac{m_{\phi_i}^4}{m_Z^4} \right)
                      \right] \sqrt{1 - \frac{4 m_Z^2}{m_{\phi_i}^2}}, 
\nonumber \\
 \Gamma(\phi_i \to ff) &=& \frac{C^{\rm color}}{8\pi} C_{\phi_i ff}^2 m_{\phi_i} \left(1-\frac{4 m_f^2}{m_{\phi_i}^2}\right)^{3/2}, 
\nonumber \\
 \Gamma(\phi_i \to \tilde{G} \tilde{G}) &\approx& \frac{1}{4\pi} C_{\phi_i \psi_X\psi_X}^2 m_{\phi_i}, 
\label{eq:Width_1}
\end{eqnarray}
where $C^{\rm color}$ is 3 (1) for squark (slepton). 
The partial width for decay to scalars is given by  
%--------------------------->
\begin{eqnarray}
 \Gamma(\phi_i \to \phi_j \phi_j) &=& \frac{1}{8\pi} \tilde{C}_{\phi_i \phi_j \phi_j}^2 
     \frac{1}{ m_{\phi_i} } 
     \sqrt{1 - \frac{4 m_{\phi_j}^2}{m_{\phi_i}^2}},
\nonumber \\
 \Gamma(\phi_3 \to \phi_1 \phi_2) &=& \frac{1}{16\pi} \tilde{C}_{\phi_3 \phi_1 \phi_2}^2 
     \frac{1}{ m_{\phi_3} } 
     \sqrt{1 - 2 \frac{ m_{\phi_1}^2 + m_{\phi_2}^2 }{m_{\phi_3}^2} 
             + \frac{ ( m_{\phi_1}^2 - m_{\phi_2}^2 )^2 }{m_{\phi_3}^4}},
  \label{eq:Width_higgs}
\end{eqnarray}
%---------------------------<
where $\tilde{C}_{\phi_i \phi_j \phi_j} = C_{\phi_i \phi_j \phi_j} + C_{\phi_j \phi_i \phi_j} + C_{\phi_j \phi_j \phi_i}$ and 
$\tilde{C}_{\phi_3 \phi_1 \phi_2} = 
  C_{\phi_1 \phi_2　\phi_3} 
+ C_{\phi_1 \phi_3　\phi_2} 
+ C_{\phi_2 \phi_1　\phi_3} 
+ C_{\phi_2 \phi_3　\phi_1} 
+ C_{\phi_3 \phi_1　\phi_2} 
+ C_{\phi_3 \phi_2　\phi_1}$. 
We can write the partial width for sgoldstino decays to several SUSY particle final states as 
%--------------------------->
\begin{eqnarray}
 \Gamma(\phi_i \to \psi_{\tilde{\chi}_j} \psi_{\tilde{\chi}_k}) &=& 
C^{\rm sym}_{jk} \frac{1}{8 \pi} m_{\phi_i} 
 \sqrt{1 - 2 \frac{m_j^2 + m_k^2}{m_{\phi_i}^2} + \frac{(m_j^2 - m_k^2)^2}{m_{\phi_i}^4}}
\nonumber \\ 
&& \times
\left[ 
 \left(C^S_{\phi_i \psi_{\tilde{\chi}_j} \psi_{\tilde{\chi}_k}} 
      +C^S_{\phi_i \psi_{\tilde{\chi}_k} \psi_{\tilde{\chi}_j}} 
      +C^{KS}_{\phi_i \psi_{\tilde{\chi}_j} \psi_{\tilde{\chi}_k}} \frac{m_j+m_k}{2} \right)^2 
      \left\{1 - \left(\frac{m_j+m_k}{m_{\phi_i}}\right)^2 \right\}
\right . \nonumber \\ 
&& \left . \quad
+\left(C^P_{\phi_i \psi_{\tilde{\chi}_j} \psi_{\tilde{\chi}_k}} 
      +C^P_{\phi_i \psi_{\tilde{\chi}_k} \psi_{\tilde{\chi}_j}} 
      +C^{KP}_{\phi_i \psi_{\tilde{\chi}_j} \psi_{\tilde{\chi}_k}} \frac{m_k-m_j}{2}\right)^2 
      \left\{1 - \left(\frac{m_j-m_k}{m_{\phi_i}}\right)^2 \right\}
\right], 
\nonumber \\
%=====
%=====
%=====
 \Gamma(\phi_i \to \psi_{\tilde{\chi}_j}^+ \psi_{\tilde{\chi}_k}^-) &=& 
\frac{1}{8\pi} m_{\phi_i}
 \sqrt{1 - 2 \frac{m_j^2 + m_k^2}{m_{\phi_i}^2} + \frac{(m_j^2 - m_k^2)^2}{m_{\phi_i}^4}}
\nonumber \\ 
&& \times
\left[
 \left(C^S_{\phi_i \psi_{\tilde{\chi}_j}^+ \psi_{\tilde{\chi}_k}^-} 
      +C^{KS}_{\phi_i \psi_{\tilde{\chi}_j}^+ \psi_{\tilde{\chi}_k}^-} \frac{m_j +m_k}{2}\right)^2 
    \left\{1 - \left(\frac{m_j +m_k}{m_{\phi_i}}\right)^2 \right\}
\right . \nonumber \\ 
&& \left . \quad 
+\left(C^P_{\phi_i \psi_{\tilde{\chi}_j}^+ \psi_{\tilde{\chi}_k}^-} 
      +C^{KP}_{\phi_i \psi_{\tilde{\chi}_j}^+ \psi_{\tilde{\chi}_k}^-} \frac{m_j -m_k}{2}\right)^2 
    \left\{1 - \left(\frac{m_j -m_k}{m_{\phi_i}}\right)^2 \right\}
\right], 
\nonumber \\
%=====
%=====
%=====
 \Gamma(\phi_i \to \tilde{f}_1^* \tilde{f}_1) &=& 
     \frac{C^{\rm color}}{16\pi} C_{\phi_i \tilde{f}_1\tilde{f}_1}^2 \frac{1}{ m_{\phi_i} } 
     \sqrt{1 - \frac{4 m_{\tilde{f}_1}^2}{m_{\phi_i}^2}},
\nonumber \\
 \Gamma(\phi_i \to \tilde{f}_1^* \tilde{f}_2) &=& 
     \frac{C^{\rm color}}{16\pi} C_{\phi_i \tilde{f}_1\tilde{f}_2}^2 \frac{1}{ m_{\phi_i} } 
     \sqrt{1 - 2 \frac{ m_{\tilde{f}_1}^2 + m_{\tilde{f}_2}^2 }{m_{\phi_i}^2} 
             + \frac{ ( m_{\tilde{f}_1}^2 - m_{\tilde{f}_2}^2 )^2 }{m_{\phi_i}^4}},
\nonumber \\
 \Gamma(\phi_i \to \tilde{f}_1 \tilde{f}_2^*) &=& \Gamma(\phi_i \to \tilde{f}_1^* \tilde{f}_2), 
  \label{eq:Width_susy}
\end{eqnarray}
%---------------------------<
where $C^{\rm sym}_{jk} = 1/2 ~(1)$ if $j=k ~(j \ne k)$. $C^{\rm color}$ is 3 (1) for squark (slepton).

The pseudo-sgoldsino decay widths are 
%--------------------------->
\begin{eqnarray}
 \Gamma(a_i \to gg) &=& \frac{2}{\pi} C_{a_i gg}^2 m_{a_i}^3,
\nonumber \\
 \Gamma(a_i \to \gamma \gamma) &=& \frac{1}{4\pi} C_{a_i \gamma\gamma}^2 m_{a_i}^3,
\nonumber \\
 \Gamma(a_i \to \gamma Z) &=& \frac{1}{8\pi} C_{a_i \gamma Z}^2 m_{a_i}^3 \left( 1-\frac{m_Z^2}{m_{a_i}^2} \right)^3,
\nonumber \\
 \Gamma(a_i \to WW) &=& \frac{1}{8\pi} C_{a_i WW_T}^2 m_{a_i}^3 \left( 1-\frac{4 m_W^2}{m_{a_i}^2} \right)^{5/2},
\nonumber \\
 \Gamma(a_i \to ZZ) &=& \frac{1}{4\pi} C_{a_i ZZ_T}^2 m_{a_i}^3 \left( 1-\frac{4 m_Z^2}{m_{a_i}^2} \right)^{5/2}, 
  \label{eq:Width_a1}
\end{eqnarray}
%---------------------------<
%
%--------------------------->
\begin{eqnarray}
 \Gamma(a_i \to ff) &=& \frac{C^{\rm color}}{8\pi} C_{a_i ff}^2 m_{a_i} \sqrt{1-\frac{4 m_f^2}{m_{a_i}^2}}, 
\nonumber \\
 \Gamma(a_i \to \tilde{G} \tilde{G}) &\approx& \frac{1}{4\pi} C_{a_i \psi_X\psi_X}^2 m_{\phi_i}, 
\nonumber \\
 \Gamma(a_2 \to a_1 \phi_i) &=& \frac{1}{16\pi} \tilde{C}_{\phi_i a_1 a_2}^2 
     \frac{1}{ m_{a_2} } 
     \sqrt{1 - 2 \frac{ m_{a_1}^2 + m_{\phi_i}^2 }{m_{a_2}^2}, 
             + \frac{ ( m_{a_1}^2 - m_{\phi_i}^2 )^2 }{m_{a_2}^4}},
\nonumber \\
 \Gamma(a_i \to \psi_{\tilde{\chi}_j} \psi_{\tilde{\chi}_k}) &=& 
C^{\rm sym}_{jk} \frac{1}{8\pi} m_{a_i}
 \sqrt{1 - 2 \frac{m_j^2 + m_k^2}{m_{a_i}^2} + \frac{(m_j^2 - m_k^2)^2}{m_{a_i}^4}}
\nonumber \\ 
&& \times
\left[
 \left(C^S_{a_i \psi_{\tilde{\chi}_j} \psi_{\tilde{\chi}_k}} 
      +C^S_{a_i \psi_{\tilde{\chi}_k} \psi_{\tilde{\chi}_j}} 
      +C^{KS}_{a_i \psi_{\tilde{\chi}_j} \psi_{\tilde{\chi}_k}} \frac{m_k - m_j}{2}\right)^2 
     \left\{1 - \left(\frac{m_j + m_k}{m_{a_i}}\right)^2 \right\}
\right . \nonumber \\ 
&& \left . \quad 
+\left(C^P_{a_i \psi_{\tilde{\chi}_j} \psi_{\tilde{\chi}_k}} 
      +C^P_{a_i \psi_{\tilde{\chi}_k} \psi_{\tilde{\chi}_j}}
      +C^{KP}_{a_i \psi_{\tilde{\chi}_j} \psi_{\tilde{\chi}_k}} \frac{m_j + m_k}{2}\right)^2
    \left\{1 - \left(\frac{m_j-m_k}{m_{a_i}}\right)^2 \right\}
\right], 
\nonumber \\
%=====
%=====
%=====
 \Gamma(a_i \to \psi_{\tilde{\chi}_j}^+ \psi_{\tilde{\chi}_k}^-) &=& 
\frac{1}{8\pi} m_{a_i}
 \sqrt{1 - 2 \frac{m_j^2 + m_k^2}{m_{a_i}^2} + \frac{(m_j^2 - m_k^2)^2}{m_{a_i}^4}}
\nonumber \\ 
&& \times
\left[
 \left(C^S_{a_i \psi_{\tilde{\chi}_j}^+ \psi_{\tilde{\chi}_k}^-} 
      +C^{KS}_{a_i \psi_{\tilde{\chi}_j}^+ \psi_{\tilde{\chi}_k}^-} \frac{m_j - m_k}{2} \right)^2 
    \left\{1 - \left(\frac{m_j +m_k}{m_{a_i}}\right)^2 \right\}
\right . \nonumber \\ 
&& \left . \quad 
+\left(C^P_{a_i \psi_{\tilde{\chi}_j}^+ \psi_{\tilde{\chi}_k}^-} 
      +C^{KP}_{a_i \psi_{\tilde{\chi}_j}^+ \psi_{\tilde{\chi}_k}^-} \frac{m_j + m_k}{2} \right)^2 
    \left\{1 - \left(\frac{m_j -m_k}{m_{a_i}}\right)^2 \right\}
\right], 
% \nonumber \\ 
%\lambda(m_i,m_j,m_k)&=& \left(1 - 2 \frac{m_j^2 + m_k^2}{m_i^2} + \frac{(m_j^2 - m_k^2)^2}{m_i^2}\right)^{1/2},
\nonumber \\
 \Gamma(a_i \to \tilde{f}_1^* \tilde{f}_2) &=& 
     \frac{C^{\rm color}}{16\pi} C_{a_i \tilde{f}_1\tilde{f}_2}^2 \frac{1}{ m_{a_i} } 
     \sqrt{1 - 2 \frac{ m_{\tilde{f}_1}^2 + m_{\tilde{f}_2}^2 }{m_{a_i}^2} 
             + \frac{ ( m_{\tilde{f}_1}^2 - m_{\tilde{f}_2}^2 )^2 }{m_{a_i}^4}},
\nonumber \\
 \Gamma(a_i \to \tilde{f}_1 \tilde{f}_2^*) &=& \Gamma(a_i \to \tilde{f}_1^* \tilde{f}_2), 
\label{eq:Width_a3}
\end{eqnarray}
%---------------------------<
where $\tilde{C}_{\phi_i a_1 a_2} = C_{\phi_i a_1 a_2}+C_{\phi_i a_2 a_1}$.

\section*{Appendix D: Higgs potential up to $O(1/F^2)$} \label{ap:D}

In this Appendix, we suppose the following lagrangian, 
%--------------------------->
\begin{eqnarray}
 \mathcal{L}_K &=& 
\int d\theta^4 \left[ 
  \left( 1 - \frac{m_{\tilde{f}_i}^2}{F^2} X^\dagger X \right) \Phi_i^\dagger e^V \Phi_i
 +\left( 1 - \frac{m_{H_{u,d}}^2}{F^2} X^\dagger X \right) H_{u,d}^\dagger e^V H_{u,d}
\right . \nonumber \\ 
&& \left . \qquad \qquad + \left\{
 -\left( \frac{\mu_k}{F} X^\dagger + \frac{B_{\mu k}}{F^2} X^\dagger X \right) H_d \cdot H_u
   + h.c. \right\} \right],
  \label{eq:Lagrangian0_A}
\\
 \mathcal{L}_W &=& 
\int d\theta^2 \left[ 
    \frac{1}{4} \left( 1 + \frac{2 M_a}{F} X \right) {\bf Tr}[W^{a \alpha} W^a_\alpha] 
  + \left( \mu_w + \frac{B_{\mu w}}{F} X \right) H_d \cdot H_u
  + \frac{A_X}{F} X X H_d \cdot H_u 
\right . \nonumber \\ 
&& \left . \qquad \qquad 
  + \left( y_e + \frac{A_e}{F} X \right) H_d \cdot L E^c 
  + \left( y_d + \frac{A_d}{F} X \right) H_d \cdot Q D^c 
  + \left( y_u + \frac{A_u}{F} X \right) H_u \cdot Q U^c 
              \right]
  + h.c. . 
\nonumber 
\end{eqnarray}
%---------------------------<
%
The D- and F-term contributions, $V_D$ and $V_F$ to the Higgs-sgoldstino potential are written as 
%--------------------------->
\begin{eqnarray}
 V_D &=& \frac{g^{\prime 2}}{8} \left( 1 + \frac{2 M_1}{F} \frac{\phi_X + \phi_X^*}{2} \right)^{-1}
         \left\{ \left( 1 - \frac{m_{H_u}^2}{F^2} |\phi_X|^2 \right) |H_u|^2
                -\left( 1 - \frac{m_{H_d}^2}{F^2} |\phi_X|^2 \right) |H_d|^2 \right\}^2 
\\ &&   
        +\frac{g_2^2}{8} \left( 1 + \frac{2 M_2}{F} \frac{\phi_X + \phi_X^*}{2} \right)^{-1}
         \left\{ \left( 1 - \frac{m_{H_u}^2}{F^2} |\phi_X|^2 \right) H_u^\dagger \sigma^i H_u
                +\left( 1 - \frac{m_{H_d}^2}{F^2} |\phi_X|^2 \right) H_d^\dagger \sigma^i H_d
        \right\}^2, 
\nonumber 
  \label{eq:VD_A}
\end{eqnarray}
%---------------------------<
% and 
%--------------------------->
\begin{eqnarray}
 V_F &=& \left( 1 - \frac{m_{H_u}^2}{F^2} |\phi_X|^2 \right)^{-1} 
         \left|- \left( \mu_{\rm eff} + \frac{B_{\mu {\rm eff}}}{F} \phi_X 
                                      - \frac{A_X}{F} \phi_X^2\right) \epsilon_{ij}H_d^i 
               +\frac{m_{H_u}^2}{F} \phi_X H_u^{*j}
               -(O_{1/F^2}) \mu_k \epsilon_{ij}H_d^i
         \right|^2
\nonumber \\ && 
        +\left( 1 - \frac{m_{H_d}^2}{F^2} |\phi_X|^2 \right)^{-1}
         \left|- \left( \mu_{\rm eff} + \frac{B_{\mu {\rm eff}}}{F} \phi_X 
                                      - \frac{A_X}{F} \phi_X^2\right)\epsilon_{ij} H_u^j 
               +\frac{m_{H_d}^2}{F} \phi_X H_d^{*i}
               -(O_{1/F^2})\mu_k \epsilon_{ij} H_u^j
         \right|^2
\nonumber \\ && 
        +\left[ 1 - \frac{m_X^2}{F^2}|\phi_X|^2
                  - \frac{m_{H_u}^2}{F^2} |H_u|^2 
                  - \frac{m_{H_d}^2}{F^2} |H_d|^2 
                  - \frac{B_{\mu k}}{F^2} \left\{ H_d \cdot H_u + (H_d \cdot H_u)^\dagger \right\}
\right]^{-1} 
\nonumber \\ \qquad &&
\times   \left|- F -2\frac{A_X}{F} \phi_X H_d \cdot H_u
                   - \frac{B_{\mu w}}{F} H_d \cdot H_u 
         \right|^2, 
  \label{eq:VF_A}
\end{eqnarray}
%---------------------------<
%
%--------------------------->
\begin{eqnarray}
O_{1/F^2} &=&  \frac{m_X^2}{F^2} |\phi_X|^2 
                       +\frac{\mu_w \mu_k + \mu_k^2 + m_{H_u}^2}{F^2} |H_u|^2
                       +\frac{\mu_w \mu_k + \mu_k^2 + m_{H_d}^2}{F^2} |H_d|^2 
\nonumber \\ && 
                       -\frac{-2A_X \phi_X - B_{\mu w} - B_{\mu k}}{F^2} H_d \cdot H_u 
                       +\frac{B_{\mu k}}{F^2} (H_d \cdot H_u)^\dagger, 
  \label{eq:VF_A_add}
\end{eqnarray}
%---------------------------<
respectively. Here, $\mu_{\rm eff} = \mu_w + \mu_k$ and $B_{\mu {\rm eff}} = B_{\mu w} + B_{\mu k}$.
%

%%%%%%%%%%%%%%%%%%%%%%%%%%%%%%%%%%%%%%%%%%%%%%%%%%%%%%%%%%%%%%%%%%%%%%%%%%%%%%%%%
% References 
%%%%%%%%%%%%%%%%%%%%%%%%%%%%%%%%%%%%%%%%%%%%%%%%%%%%%%%%%%%%%%%%%%%%%%%%%%%%%%%%%

\end{document}